\documentclass[namedreferences]{solarphysics}

\usepackage[hyperref,optionalrh]{spr-sola-addons} 
\usepackage{acronym}
\usepackage{booktabs}
\usepackage{longtable}
\usepackage{pdflscape}
\usepackage{adjustbox}
\usepackage{lscape}
\usepackage{rotating}
\usepackage[width=1.35\textwidth]{caption}
\usepackage{xcolor}
\usepackage{amsopn}

\renewcommand{\thetable}{\arabic{table}}
\DeclareMathOperator{\atantwo}{atan2}

\acrodef{FF}{fast forward}
\acrodef{FR}{fast reverse}

\renewcommand{\vec}[1]{{\mathbfit #1}}

\newcommand{\kms}{km s$^{-1}$} 

\newcommand{\dV}{\Delta V}
\newcommand{\MA}{M_{\mathrm A}}

\begin{document}

\begin{article}
\begin{opening}

\title{Statistical Analysis of Interplanetary Shocks from Mercury to Jupiter}


\author[addressref={aff0},corref,email={cpereza.art@igeofisica.unam.mx}]{\inits{C.}\fnm{Carlos}~\lnm{Pérez-Alanis}}

\author[addressref={aff1},email={miho.janvier@universite-paris-saclay.fr}]{\inits{M.}\fnm{Miho}~\lnm{Janvier}}

\author[addressref={aff2},email={Teresa.Nieves@nasa.gov}]{\inits{T.}\fnm{Teresa}~\lnm{Nieves-Chinchilla}}

\author[addressref={aff4},email={ernesto@igeofisica.unam.mx}]{\inits{E.}\fnm{Ernesto}~\lnm{Aguilar-Rodr\'{i}guez}}

\author[addressref={aff5,aff6},email={pascal.Demoulin@obspm.fr}]{\inits{P.}\fnm{Pascal}~\lnm{D\'{e}moulin}}

\author[addressref={aff7,aff8,aff9},email={p.coronaromero@igeofisica.unam.mx }]{\inits{P.}\fnm{Pedro}~\lnm{Corona-Romero}}

\address[id=aff0]{Posgrado en Ciencias de la Tierra, Universidad Nacional Aut\'{o}noma de M\'{e}xico, UNAM, CDMX, M\'{e}xico.}

\address[id=aff1]{Université Paris-Saclay, CNRS, Institut d’Astrophysique Spatiale, Orsay, France.}

\address[id=aff2]{Heliophysics Science Division. NASA Goddard Space Flight Center, Greenbelt, MD, USA.}

\address[id=aff4]{Instituto de Geof\'{i}sica, Unidad Michoac\'{a}n, Universidad Nacional Aut\'{o}noma de M\'{e}xico, UNAM, Morelia, M\'{e}xico.}

\address[id=aff5]{LESIA, Observatoire de Paris, Universit\'{e} PSL, CNRS, Sorbonne Universit\'{e}, Universit\'{e} Paris Cit\'{e}, 5 place Jules Janssen, 92195 Meudon, France.}

\address[id=aff6]{Laboratoire Cogitamus, 75005 Paris.}

\address[id=aff7]{Cátedras-CONACyT, Instituto de Geofísica Unidad Michoacán, Universidad Nacional Autónoma de México, 58190 Morelia, Michoacán, México.}

\address[id=aff8]{Space Weather National Laboratory (LANCE), Instituto de Geof\'{i}sica, Unidad Michoac\'{a}n, Universidad Nacional Aut\'{o}noma de M\'{e}xico, 58190 Morelia, Michoacán, Mexico.}

\address[id=aff9]{Mexican Space Weather Service (SCIESMEX), Instituto de Geof\'{i}sica, Unidad Michoac\'{a}n, Universidad Nacional Aut\'{o}noma de M\'{e}xico, 58190 Morelia, Michoacán.}

\runningauthor{Pérez-Alanis \textit{et al.}}

\runningtitle{Statistical analysis of interplanetary shocks from Mercury to Jupiter}


\begin{abstract}
In situ observations of interplanetary (IP) coronal mass ejections (ICMEs) and IP shocks are important to study as they are the main components of the solar activity. Hundreds of IP shocks have been detected by various space missions at different times and heliocentric distances. Some of these are followed by clearly identified drivers, while some others are not. 
In this study, we carry out a statistical analysis of the distributions of plasma and magnetic parameters of the IP shocks recorded at various distances to the Sun. We classify the shocks according to the heliocentric distance, namely from 0.29 to 0.99 AU (Helios-1/2); near 1 AU (Wind, ACE and STEREO-A/B); and from 1.35 to 5.4 AU (Ulysses). We also differentiate the IP shocks into two populations, those with a detected ICME and those without one. We find, as expected, that there are no significant differences in the results from spacecraft positioned at 1 AU. 
Moreover, the distributions of shock parameters, as well as the shock normal have no significant variations with the heliocentric distance. Additionally, we investigate how the number of shocks associated to stream-interaction regions (SIRs) increases with distance in proportion of ICME/shocks.  From 1 to 5 AU, SIRs/ shock occurrence increases slightly from 21\% to 34\%, in contrast ICME/shocks occurrence decreases from 47\% to 17\%. 
We find also indication of an asymmetry induced by the Parker spiral for SIRs and none for ICMEs. 
\end{abstract}

\end{opening}

\section{Introduction}
\label{Sec:introduction}

Since the beginning of the space age several space missions have carried out in situ measurements of the properties of the heliosphere. Space missions such as Ulysses, focused on studying the solar wind from high solar latitudes, and different missions located at 1 AU, such as ACE, WIND or the twin STEREO spacecraft provide us with better in situ observations to improve our understanding of solar phenomena through the interplanetary (IP) medium. Furthermore, new solar missions such as the Parker Solar Probe (PSP) spacecraft and Solar Orbiter are currently improving the understanding of solar processes and our knowledge of solar wind at different locations in the solar system.

The solar wind is characterized by a decrease in its density while we move outward from the Sun into the IP medium. The solar wind does not smoothly flow out, and is the place where transients such as shocks \citep{cane1985evolution,cane1987energetic}, coronal mass ejections \citep[CMEs;][]{gopalswamy2006properties} and stream interaction regions (SIRs), between slow and fast solar winds \citep{jian2009multi} occur. Shocks take place when a wave experiences a steepening process in a collisionless plasma and, the mean free-path of particles is around 1 AU \citep{1991pspa.book.....P}. When a wave that moves through the magnetized plasma travels faster than the solar-wind Alfv\'{e}n speed, the amplitude of this wave increases rapidly and the nonlinear effects become important, so a shock wave eventually occurs \citep{burgess2015collisionless}. The sudden transition between supersonic and subsonic flows across a shock is characterized by an abrupt change in pressure, temperature, density, and magnetic field intensity in the medium \citep{burlaga1971reverse,eselevich1982shock,sagdeev1991collisionless}. In the IP medium and close to the Sun, shocks are mainly caused by IP CMEs (ICMEs), when they propagate faster than the fast mode wave speed with respect to the ambient solar wind, and expanding through the IP medium \citep{gosling1968satellite}. The solar-wind material behind the shock is named an ICME sheath. Behind the ICME sheath, the region where there are less intense magnetic fluctuations than in the sheath is known as the magnetic ejecta 
\citep{winslow2015interplanetary}. Also, in other cases, shocks can also be generated by interaction between high-and slow-speed solar streams which form SIRs \citep{gosling1972compressions,gosling1976solar,richardson2018solar}.

Coronagraphic and heliospheric observations indicate that IP shocks propagate through the solar wind along a broad and roughly spherical front, ahead of plasma and magnetic field ejected from the CME \citep{hundhausen2012coronal}. In general, the geometry of the whole ICME/shock structure can be identified as a bubble expanding as it moves away from the solar source \citep{berdichevsky2000interplanetary,berdichevsky2003geometric}. However, the true three-dimensional geometry or morphology of the ICME/shock structure has remained not well constrained due to the limitations to identify and track them using the traditional image processing technique \citep{howard2006tracking,ontiveros2009quantitative,thernisien2011implementation}.
Several authors have studied the morphology and propagation of an IP shock through the bow shock and magnetosheath using numerical magnetohydrodynamic (MHD) simulations \citep{koval2005deformation,samsonov2006numerical}. On the other hand, since measurements detected by in situ spacecraft only provide a single observation point (apart from rare events of longitudinally aligned spacecraft, i.e. during the early mission years of the twin STEREO spacecraft), the in situ data of the ICME/shock properties are considered local and the obtained results are therefore limited. 

In recent years, some studies analyzed the shock properties as a function of their distances to the shock apex. Using a sample of in situ observations by the Ulysses mission, \citep{hoang1995interplanetary} investigated the correlation of the shock strength with the heliocentric distance. This kind of study was extended to shocks detected from 0.28 to 1 AU by \cite{lai2012radial}. They concluded that there was no correlation between the shock magnetic-field enhancement and the radial distance. Finally, \cite{janvier2014mean} and \cite{janvier2015comparing}, from the statistical analysis of a large set of IP shocks observed at 1 AU, found that there is no privileged direction of the shock normal vector around the Sun apex line. 

In this study, we aim to understand the spatial and temporal variation of the IP-shock structures throughout different locations in the inner heliosphere. The paper is structured as follows: Section 2 presents the catalogs of IP shocks that we use to form our set of events to be analyzed throughout this study. In Section 3 we investigate at 1 AU the distributions of the upstream/downstream parameters of shocks as well the correlation between the location angle $\lambda$ (the angle between the shock normal and the radial direction), which allows us to define the position of the shock crossing of the spacecraft with respect to the shock front for a given shock shape. We extend this study for different heliospheric distances in Section 4 
and we explore a possible asymmetry between the shape of the shock-driver with the east/west direction of the structure as well. We discuss and conclude the study in Section 5.

\section{IP Shocks Selection: Catalogs Covering Heliospheric Distances 0.3 to 5.4 AU.}
\label{Sec:selection}

\subsection{Missions Included in the Present Study} \label{Sec:missions}

Several authors have previously studied IP shocks using in situ observations. The duration of some space missions have allowed to report all these IP shocks in catalogs of events spanning a few years up to several solar cycles. Based on these previous studies, we made an inspection throughout different websites and research papers in order to obtain IP shocks and, when available, ICMEs associated with IP shocks detected at $1$ AU, and completed the dataset with IP shocks observed by Helios-1/2 as well as Ulysses. 

The two Helios-1/2 space missions were launched in 1974 (Helios-1) and 1976 (Helios-2), and covered the heliospheric distance range between 0.29 and 0.99 AU from the Sun. The main purpose of the missions was to make pioneering measurements of the IP medium from the vicinity of the Earth to near the Sun. Measurements of the IP medium were made with the Helios magnetic-field and plasma experiments, described in \cite{neubauer1976initial} and \cite{rosenbauer1977survey}.

For the data at 1 AU, we used a combination of data from the Wind, ACE, and STEREO A-B missions. The Wind spacecraft was launched on 1 November 1994, with the aim to study the IP medium and its effects on the Earth's magnetosphere. After several orbits through the magnetosphere, the Wind spacecraft was placed in a halo orbit around the L1 Lagrange point, where it currently remains and constantly records measurements of the solar wind from its instruments on board \citep{ogilvie1997wind}. The main instruments detecting shocks are the Magnetic Field Investigation  \citep[MFI,][]{lepping1995wind}, the Solar Wind Experiment  \citep[SWE,][]{ogilvie1995swe} and the Radio and Plasma Wave Investigation \citep[WAVES,][]{bougeret1995waves}. 

The Advanced Composition Explorer (ACE) space mission was designed to study space-borne energetic particles and solar-weather monitoring from the Sun-Earth L1 Lagrange point. Since its launch in August 25, 1997, it has continued to provide near-real-time coverage of solar-wind parameters and to measure solar energetic particles intensities. For the shock determination, magnetic-field and plasma parameters are given by the Magnetic Field Instrument (MAG) and the Solar Wind Electron, Proton and Alpha Monitor \citep[SWEPAM,][]{smith1998ace}.

The Solar Terrestrial Relations Observatory (STEREO) is a solar space mission, with two identical spacecraft that were launched on October 26, 2006. Once in heliospheric orbit, STEREO-B trails the Earth (at 1.05 AU) while STEREO-A leads it (at 0.95 AU). As viewed from the Sun, the two spacecraft separate at approximately 44 to 45 degrees per year \citep{kaiser2008stereo}. The data are provided by the instruments IMPACT \citep{acuna2008stereo,luhmann2008stereo} and PLASTIC \citep{galvin2008plasma}. 

Finally, we also considered data from the Ulysses mission \citep{wenzel1992ulysses}. This mission was unique in the history of the exploration of our solar system: launched in October 1990 to explore the heliosphere within a few astronomical units of the Sun over a full range of heliographic latitudes, passing over the south pole of the Sun in mid-1994 and over the north pole in mid-1995. Advanced scientific instrumentation carried on board the spacecraft provide a comprehensive set of observations to study the solar wind at all latitudes. Among the main instruments, the data for the shocks come from the magnetic fields experiment \citep[VHM/FGM,][]{balogh1992magnetic} and solar wind-plasma experiment \citep[SWOOPS,][]{bame1992ulysses}.

\subsection{Interplanetary Shocks, ICMEs, and SIRs Databases}
\label{Sec:databases}

\subsubsection{Interplanetary Shock Database}

While several papers have reported different catalogs for IP shocks detected by the space missions described above, it is also important that the shock analysis is consistent from one catalog to another. On one hand, there are numerous techniques aimed at determining shock normals, shock speeds, values of upstream/downstream plasma, and field parameters, such as magnetic/velocity coplanarity or mixed mode method \citep{paschmann1998analysis}. On the other hand, the detection of IP shocks depends on the temporal accuracy of the measurements, as well as a definition of what is considered as the upstream and downstream intervals of the shock. Therefore, it is necessary that the databases are consistent from one catalog to another to provide well-documented shocks for future investigations and to evaluate the accuracy of several shock normal determination techniques \citep{russell1983multiple}.

In our study, we chose the comprehensive Heliospheric Shock Database (HSD) (www.ipshocks.fi/), which provides shock properties detected by different missions \citep{kilpua2015properties}. This database was compared with the Harvard Smithsonian Center of Astrophysics Wind shock database (CfA Interplanetary Shock Database), which can be found at lweb.cfa.harvard.edu$\slash$shocks$\slash$. The latter also provides detailed analyses of IP shocks observed at 1 AU by Wind and ACE spacecraft. Since each database offers different methods to identify and characterize the observed shocks, we compared in Appendix \ref{Appendix:A} both databases, through the correlation of the shock normal values, derived by different methods, finding that the mixed-mode method offers the best correlations. We found a good agreement between the two databases, which gives us the confidence to use the data within HSD in the following. 

In this database, the shocks are identified using two techniques: (1) through visual inspection of the solar-wind plasma and magnetic-field observations, and (2) with an automated shock detection algorithm (see the documentation available on the website, and references therein). For each spacecraft, the parameters available in the database are: the IP magnetic-field vector, the solar-wind velocity and the bulk speed of protons/ions, the solar-wind proton/ion number density, the solar-wind proton/ion temperature or the most probable thermal speed, the spacecraft position, and the shock type (\ac{FF} or \ac{FR}). The upstream and downstream intervals were chosen so that the mean values are taken sufficiently far from the peak of the shock, and determined over a fixed analysis interval ($\sim 8$ minutes). Additionally, the number of data points depends on the resolution of the plasma data of each spacecraft and the shock type. 

We only considered the \acs{FF} shocks that propagate away from the Sun.  These types of shocks are the most commonly observed in the solar wind up to 1 AU \citep{pitvna2021turbulence}. The \acs{FR} shocks propagate toward the Sun, but are carried outward from the Sun by the solar-wind flow. 
Additionally, we only considered shocks that satisfy the following upstream/downstream conditions: 
  (1) the magnetic field ratio $B_{d}/B_{u} \geq 1.2$, 
  (2) the proton density ratio $N_{p}^{d}/N_{p}^{u} \geq 1.2$, 
  (3) the proton temperature ratio $T_{p}^{d}/T_{p}^{u} \geq 1.2$, and (4) the magnetosonic Mach number $\MA > 1$.
  This allows us to compare our results to the statistical study of \citet{kilpua2015properties}. 
\setcounter{table}{0}

\begin{table}[!t] 
	\centering		 
		\begin{tabular}{l|cccc}
			\hline
			\multicolumn{5}{c}{\textbf{Total number of IP shocks }} \\
			\cmidrule{1-5}
			\textbf{Spacecraft} & \textbf{Time coverage}&\textbf{R [AU]}&\textbf{Source}& \textbf{No. events}\\
			Helios-1/2 & 1975-1981 &0.29-0.99& a & 103 \\
			Wind & 1994-2017 &1& a & 469 \\
			ACE & 1998-2013 &1& a & 255 \\
			STEREO-A/B & 2007-2014 &0.95/1.05& a & 204 \\
			Ulysses & 1990-2009 &1.3-5.4& a & 238 \\
			
			\hline
			\multicolumn{5}{c}{\textbf{Number of shocks with an associated ICME }}  \\
			\cmidrule{1-5}
			Helios-1/2 & 1975-1981 &0.29-0.99& e & 30 \\
			Wind & 1994-2017 &1& b & 185 \\
			ACE & 1998-2013 &1& c & 125 \\
			STEREO-A/B & 2007-2014 &1& d & 98 \\
			Ulysses & 1990-2009 & 1.3-5.4 & f & 38 \\

			\hline
			\multicolumn{5}{c}{\textbf{Number of shocks with an associated SIR }}  \\
			\cmidrule{1-5}
			Helios-1/2 & 1975-1981 &0.29-0.99& e & 9 \\
			Wind, ACE and STEREOs  & 1994-2017 &1& g & 90 \\
			Ulysses & 1990-2009 & 1.3-5.4 & h & 80 \\

			\cmidrule{1-5}
			\multicolumn{4}{l}{$^{a}$ipshocks.fi/}  \\
			\multicolumn{4}{l}{$^{b}$\citet{nieves2018understanding}}  \\
			
			\multicolumn{4}{l}{$^{c}$\cite{regnault20}}  \\
			\multicolumn{4}{l}{$^{d}$\cite{jian2006properties,jian2013solar}   } \\
			\multicolumn{4}{l}{$^{e}$\cite{lai2012radial} }  \\
			\multicolumn{4}{l}{$^{f}$\cite{richardson2014identification} }\\
			\multicolumn{4}{l}{$^{g}$\cite{jian2006properties,jian2019solar} }\\
			\multicolumn{4}{l}{$^{h}$\cite{jian2008stream} }\\
	\end{tabular}
		\caption{Summary of the number of shocks detected at different heliocentric distances by Helios-1/2,
		Wind, ACE, STEREO-A/B, and Ulysses spacecraft, as well as the sources for the catalogs used. }
\label{table_1}
\end{table}

\subsubsection{ICME Database}

Whenever possible, we associate IP shocks with a driver, namely an ICME, by cross-referencing the list of shocks with that of reported ICMEs. Unfortunately, the number of ICMEs detected by the Helios mission being scarce for a statistical study ($30$ events), we focus here on ICME-driven shocks detected close to Earth's orbit and Ulysses.

For the near-Earth analysis, we used for the Wind data the catalog from  \citet{nieves2018understanding}, which can also be found on the Wind NASA website\footnote{wind.nasa.gov/ICMEindex.php}. In this catalog, the authors report the start and end time of the ICME and magnetic ejecta. For ACE data, we used the catalog provided by \cite{regnault20} from 1997 to 2017, which is based on the revision of the Richardson and Cane ICME list \citep{cane2003interplanetary,richardson2010near}. Finally, 
\cite{jian2006properties,jian2013solar} provide us with an ICME list observed by STEREO-A/B between 2006-2014.
For ICMEs observed by Ulysses, we used the list provided by \cite{richardson2014identification}, which covers three spacecraft orbits, during 1996 to 2009, along with the start/end time of the events, solar-wind parameters during the ICME interval, the spacecraft heliocentric distance, heliolatitude, among others.

Through an inspection of both IP shock and ICMEs databases, we associate each ICME with the closest shock from the HSD list by comparing the time difference of the shock and the disturbance of the ICME. We select events with a time difference of less than two hours between the ICME fronts  and shocks, leading to the creation of a catalog of ICME-associated IP shocks observed by all these spacecraft. 

\subsubsection{SIRs Database}

Stream interaction regions (SIRs) are characterized by a fast wind catching up with a slower one. This induces a region of compression and the total pressure ($P_{t}$) reaches a maximum inside. 
Other characteristics may also be present in SIRs, such as the compression of proton number density and of the magnetic field, as well as a significant temperature increase.  The plasma flow deflection, both in the preceding and in the following fast wind is also frequently observed.  These properties are useful for a better identification of SIRs.  
We used several catalogs of SIRs around L1: for Wind and ACE we considered the catalog developed by \citet{jian2006properties} covering the period of 1995-2004 (Wind) and 1998-2004 (ACE). For SIRs observed by STEREO-A/B we used the catalog provided by \citet{jian2019solar} between 2007-2016. In the case of the Ulysses mission, we used the list of SIR events based on studies developed by \citet{jian2008stream} from 1992 to 2005. For each catalog, the authors report the start and end time of the SIRs, 
whether a forward/reverse shock or an ICME associated with the SIR is detected, the total pressure across the discontinuity and the maximum value of the solar-wind speed, proton temperature and magnetic-field magnitude. In addition, we also use in-situ observations of plasma and magnetic field, for each mission, to identified SIRs, following the criteria mentioned above,  and complement the catalogs presented. 

The numbers of events and results from associating shocks, ICMEs, and SIRs as well as the time coverage, heliocentric distance, and the sources used in this study, are all reported in Table~\ref{table_1}.

\begin{figure} [!t]
	\begin{center} 
		\includegraphics[width=10cm]{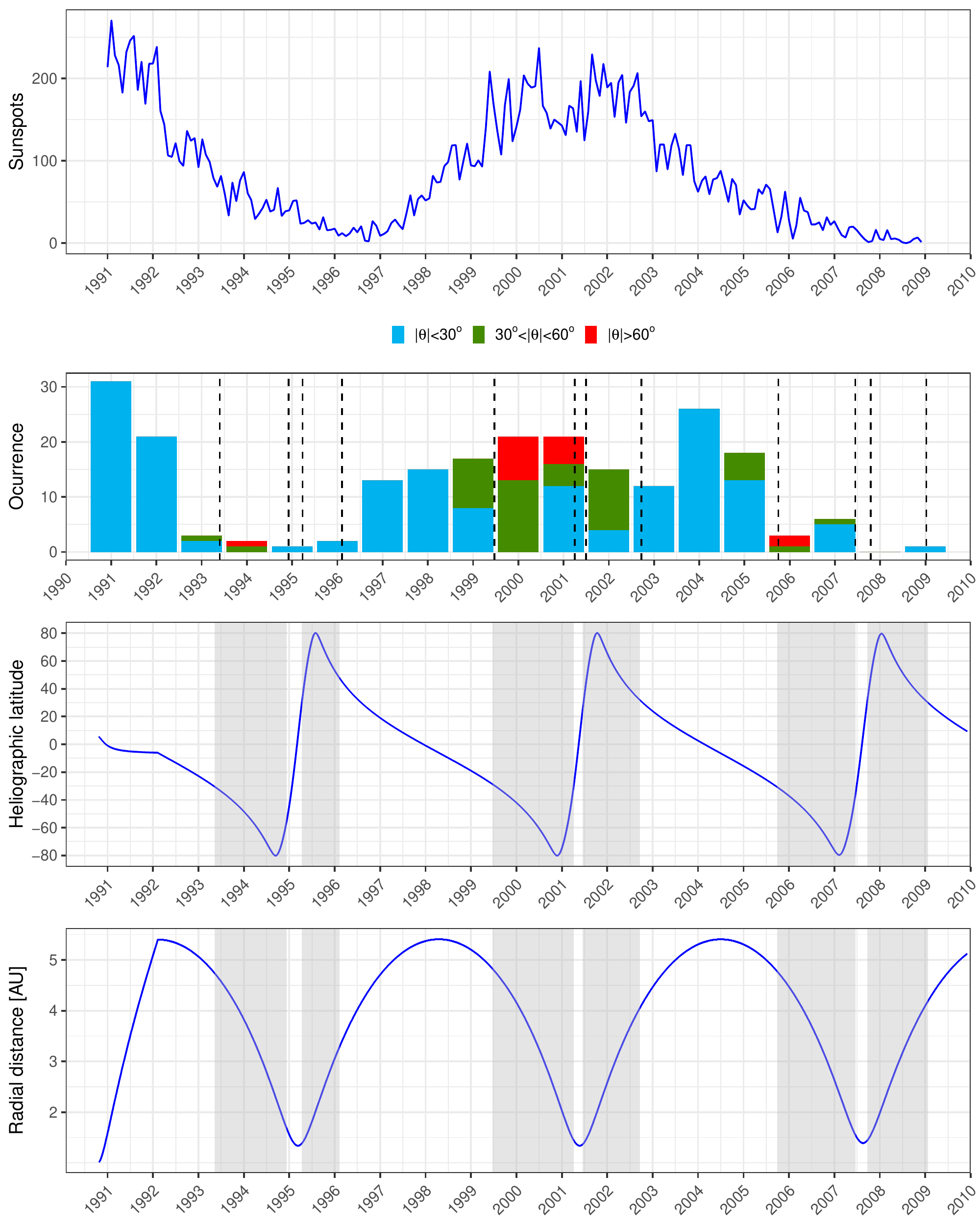}
	\end{center}
	\caption{First panel: the sunspot temporal variation between 1991 and 2009 (binned to 12 days).   Second panel: distribution of the IP shocks detected by Ulysses represented with a stacked histogram of the occurrence of the shocks for three categories of observed Heliospheric latitude $\theta$: in blue for shocks detected with $|\theta|$ below  $30^{\circ}$, in green $|\theta|$ between $30^{\circ}$ and $60^{\circ}$, and in red those observed with $|\theta|$ greater than $60^{\circ}$. The third panel shows the Ulysses heliographic latitude of Ulysses, with the shaded area showing the time intervals when Ulysses $|\theta|$ goes above $30^{\circ}$. Finally the last panel shows the radial position of Ulysses.} 
		\label{UlyssesProfiles1}
\end{figure} 

\begin{table}[!t] 
\begin{tabular}{c | c  c  c  r }
\hline
Radial distance [AU] & \multicolumn{3}{c}{Heliographic latitude [$\theta$]}  & \\
        & $|\theta| < 30^{\circ}$   & $30^{\circ} < |\theta| <60^{\circ}$  & $ |\theta| > 60^{\circ}$ & \\
\hline        
$[1,2.5]$ & 27  & 5  & 11 & \\
$[2.5,4]$ & 13  & 22 & 7 & \\
$[4,5.5]$ & 133 & 20  & 0 &  \\
Total     & 173 & 47 & 18 & \\ 
\hline
\end{tabular}
		\caption{Number of fast-forward shocks observed by Ulysses, during 1990 to 2009, organized by different radial and latitude intervals.}
\label{table_Ulysses}
\end{table}

\subsection{Latitude of Shocks Observed by Ulysses}
\label{Sec:Ulysses}

The Ulysses orbit is highly inclined at approximately 80 degrees to the heliographic equator, allowing it to scan the heliosphere at different heliographic latitudes, and in particular to pass nearly over the poles of the Sun. In order to investigate shock properties at different heliospheric distances, we first analyze how many IP shocks were observed by the spacecraft within different latitude ranges, so as to not mix the latitude and radial dependencies. 

To investigate the properties of the IP shocks detected by Ulysses and their behaviour throughout Solar Cycle 22 (SC22) and 23 (SC23), the second row of Figure \ref{UlyssesProfiles1} shows the occurrence of these events. We separate these shocks in three latitude ranges, displayed in a stacked histogram. We compare this evolution with the sunspot-number variation throughout the whole mission in the first row (data taken from the site www.swpc.noaa.gov/).  In Table \ref{table_Ulysses} we indicate the number of all shocks, during 1990 to 2009, separated within three different radial and latitude intervals.  

The occurrence of events shows a correlation with the solar activity, with a larger number of events after the maximum of the SC22 (in 1990) and SC23 (in 2000). Next, the radial and latitude position of Ulysses is changing with time (last two rows of Figure \ref{UlyssesProfiles1}). The radial evolution could affect the number of detected cases if the angular width of the shocks, as seen from the Sun, evolves with distance. Next, most of the shocks were detected at low latitudes, at less than $30^{\circ}$, and at distances greater that 4 AU, with 133 shocks observed. A possible explanation is that fast and slow winds are found at low latitude \citep{goldstein1996ulysses,neugebauer1998spatial}, creating an environment where it is easier to create a shock even with moderately fast ICMEs. These results agree with those of \cite{gonzalez1996interplanetary}, since they found a decrease in the number of forward shocks, during 1990 to 1994, at high latitudes (greater than $38^{\circ}$). In the case of CIR-forward shocks, they found that these events are stronger at low latitudes, but weaker and unlikely at high latitudes. 
We should also note that these detections are also biased by the length spent in these latitude intervals, as Ulysses spent comparatively little time at high latitudes. More precisely, Ulysses spent 3503 days orbiting at less than $30^{\circ}$ in absolute heliographic latitude, 2100 days between $30^{\circ}$ and $60^{\circ}$, and only 1208 days greater than $60^{\circ}$. Besides, the combination of Ulysses latitude and solar activity modulates the number of shocks in each year. In both sets of solar minimum, in high latitude phases Ulysses spent many months in continuous fast coronal hole solar wind, hence very few shocks were detected, in contrast to the significant number of events even at high latitudes during the solar maximum phases, when coronal streamers and regions of slow solar wind spanned all latitudes.

The event counting is also biased in the radial direction since Ulysses spent a total of 3673 days orbiting between 4 and 5.5 AU, while in the intervals [1,2.5] AU and [2.5,4] AU respectively, the spacecraft only spent 1414 and 1748 days.  Considering that most of shocks detected at low latitudes were in the interval [4,5.5] AU,  we will focus in Section \ref{Sec:distances} on this interval, which provides a large enough number of cases to investigate statistical distributions. Besides, the other intervals especially those closer to Earth are covered by the large statistics at 1 AU. 
 
\begin{figure} [!t]
	\begin{center} 
		\includegraphics[scale=0.65]{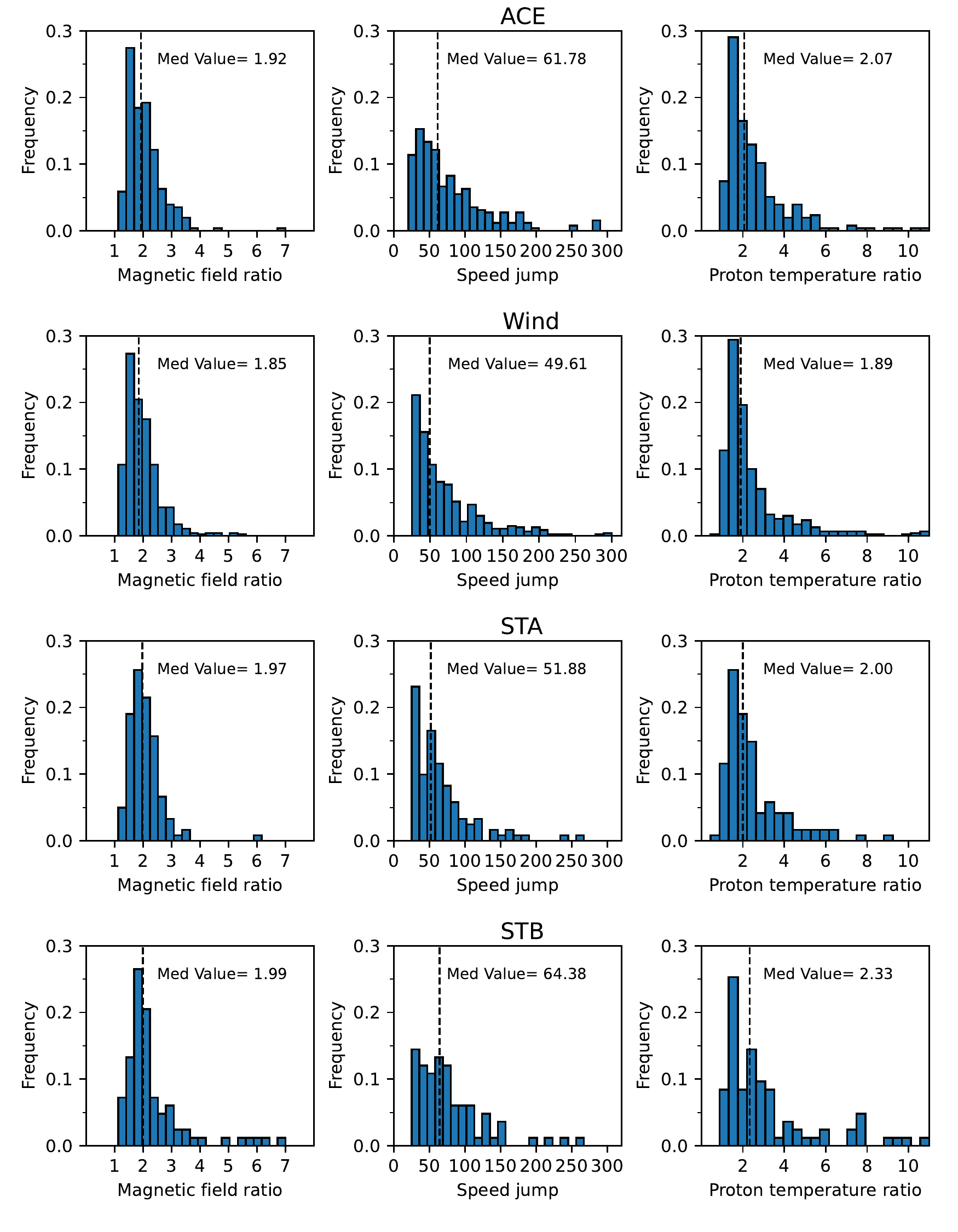}	\end{center}
	\caption{Distributions of the plasma and magnetic properties for all the IP shocks detected at 1 AU. From left to right: the magnetic-field ratio, the solar-wind speed jump and the proton-temperature ratio. The first row corresponds to shocks detected by ACE, the second row that for Wind, and the third and fourth rows the STEREO-A and B missions. The vertical dashed-line indicates the median value of each distribution. }
	\label{Allshocks1AU_HistSW}
\end{figure} 

\section{Multi-spacecraft Analysis of Interplanetary Shocks at 1 AU}
\label{Sec:1au}

In this section we focus on the results obtained for spacecraft positioned at 1 AU: ACE, Wind and STEREO-A/B. We compare the results with each other and investigate the importance of the resulting differences.
We also study the relation of the shock parameters and the location angle. While a similar study was conducted in \citet{janvier2014mean} for shocks at 1 AU, the large coverage of years and spacecraft from the HSD source allows us to look at the results in a more thorough way.

\subsection{Comparing Spacecraft Results at 1 AU}
\label{Sec:Comparing1au}
    
Figure \ref{Allshocks1AU_HistSW} shows the shock parameter distributions for all the IP shocks detected by spacecraft positioned around 1 AU, namely by ACE (first row), Wind (second row) and STEREO-A/B (third and fourth rows). We report, from left to right, the distributions of the magnetic-field ratio, the jump in the solar-wind speed and the proton-temperature ratio at the shock. All the events reported for each spacecraft have been grouped into 24 bins. All the histograms display a normalized frequency (probability), therefore allowing a direct comparison between the different rows/columns. We indicate the median value for each histogram with a vertical dashed line.
  
\begin{table}
		\begin{adjustbox}{width=1.0\textwidth}
	\begin{tabular}{cc|cccccc|cccccc}\hline	
			s/c & Number &\multicolumn{6}{c|}{Most probable value} 
			&\multicolumn{6}{c}{ Median value}  \\			&  & $B_{\rm d}/B_{\rm u}$ & $N_{\rm p}^{\rm d}/N_{\rm p}^{\rm u}$ & $T_{\rm p}^{\rm d}/T_{\rm p}^{\rm u}$ & $\MA$ & $\dV$ & $V_{\rm sh}$ &   $B_{\rm d}/B_{\rm u}$ & $N_{\rm p}^{\rm d}/N_{\rm p}^{\rm u}$ & $T_{\rm p}^{\rm d}/T_{\rm p}^{\rm u}$ & $\MA$ & $\dV$ & $V_{\rm sh}$ \\
			\hline
			
			ACE & 255 & 1.54 & 1.76&  1.54&  1.54&  41.5&  462 & 1.9& 2.0& 2.0& 1.7& 61.8& 509 \\ 
			
			
			WIND & 469 & 1.54&  1.76&  1.54&  1.26&  30.5&  462 & 1.9& 1.9& 1.9& 1.6& 49.6 & 457  \\ 
			
			
			STA & 121 &1.82&  1.76&  1.54&  1.27&  30.5&  374 & 1.9& 2.4& 2.0& 1.4& 51.9 & 408 \\ 
			
			
			STB & 83 &1.82&  1.76&  1.54&  1.26&  30.5&  418 & 1.9& 2.5& 2.3& 1.4& 64.3 & 423 \\ 
			
			All spacecraft & 928 &1.54&  1.76&  1.54&  1.26&  30.5&  418 & 1.9& 2.0& 1.9& 1.6& 55 & 457 \\ 
			
			\hline
	\end{tabular}
	\end{adjustbox}
\caption{ 
Statistical quantities from the ratio distributions of the IP shocks detected at 1 AU \\ (see Appendix \ref{Appendix:A}).
The parameters are: 
the magnetic field ratio $B_{\rm d}/B_{\rm u}$ (where the subscript d\\ is for downstream and u for upstream), 
the proton density ratio $N_{\rm p}^{\rm d}/N_{\rm p}^{\rm u}$, 
the proton temperature\\ ratio $T_{\rm p}^{\rm d}/T_{\rm p}^{\rm u} $, the magnetosonic Mach number $\MA$,
the velocity jump $\dV= V_{\rm d}-V_{\rm u}$,
and the\\ shock velocity $V_{\rm sh}$. 
The units of $\dV$ and $V_{\rm sh}$ are \kms .\\
}
\label{table_icmes_1au}
\end{table}

We find that most of the distributions for the same parameters display similar trends, as seen by the shape of the histograms (Figure \ref{Allshocks1AU_HistSW}) as well as the most probable values and medians (Table \ref{table_icmes_1au}).
The distributions for the speed present slight variations between spacecraft results, which probably are due to the statistical fluctuations. Still, the most probable values does not present important variations from one spacecraft to another. 

More precisely, the distributions of the magnetic-field ratios start with an abrupt increase to the peak and a tail extending toward $\sim$ 4. The speed jump, and the temperature ratio distribution show a similar tendency, with a shape similar to a Gamma-distribution. Quantitative results are summarized in Table~\ref{table_icmes_1au}. The standard deviations obtained with different spacecraft are similar, then we only report the standard deviation grouping all the spacecraft data:  
$\sigma_{B_{\rm d}/B_{\rm u}} =0.69$,  $\sigma_{N_{\rm p}^{\rm d}/N_{\rm p}^{\rm u}} =1$, $\sigma_{T_{\rm p}^{\rm d}/T_{\rm p}^{\rm u}} =2.62$, $\sigma_{\MA} =0.89$, $\sigma_{\dV} =51$ \kms, and $\sigma_{V_{\rm sh}} =151$ \kms . These values are lower than the median values as shown in the Table \ref{table_icmes_1au}. 

In conclusion, we do not find major differences between the 4 different spacecraft positioned at 1 AU, especially when considering the spacecraft with the highest number of detected shocks (ACE and Wind). This is expected, for ACE and Wind which are both spatially and temporally close. Within the limitation due to statistical fluctuations of STEREO results, this is extended to broader longitude and time differences.

We further investigated the evolution of the parameters with the solar cycle, and we found no significant variations with solar cycle (not shown). This is in agreement with Figure 3 of \cite{kilpua2015properties}, where the authors presented the solar cycle variations of annual medians of the shock parameter between 1993 to 2013. They found no significant variations of the shock parameters with the solar cycle, and our study confirms this. 

\begin{figure} [!t]
	\begin{center} 
		\includegraphics[scale=0.42]{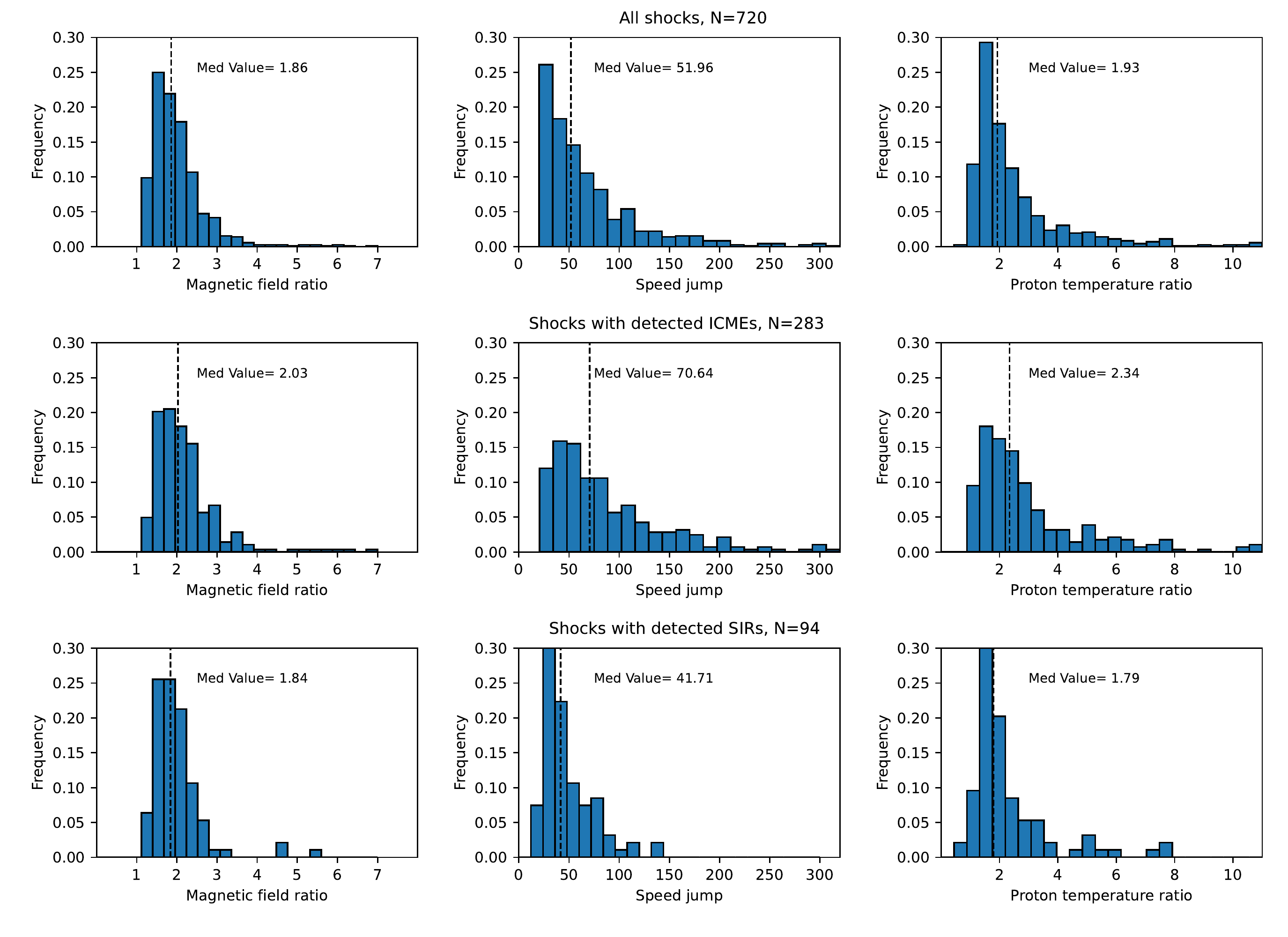}
	\end{center}
	\caption{Distributions of the IP shock ratio properties using observations at 1 AU. From left to right: magnetic-field ratio, solar-wind speed jump and proton-temperature ratio. The first row indicates all the shocks reported at 1 AU keeping only Wind data for shocks observed by several spacecraft. The second row corresponds to all shocks with a clearly identified ICME in the downstream region of the shock, and the third row shocks with SIRs detected. 
	The dashed-line indicates the median value of each distribution.}
	\label{shocks1AU_HistSW}
\end{figure} 

\subsection{Properties Distributions of IP Shocks, with and without a Detected ICME.}
\label{Sec:WithWithoutICMEs}

Having checked that there are no significant biases between different spacecraft, we next investigate whether shocks with a detected or a non-detected ICME are different at 1 AU.

Figure \ref{shocks1AU_HistSW} shows the distributions of all the shock properties shown in Figure \ref{Allshocks1AU_HistSW} for all the spacecraft considered together. Since Wind and ACE are orbiting around L1, in some cases, the same shock has been observed for both spacecraft. Therefore for these events, we might be adding the same shock twice in our statistics. 
To solve this, we compared the starting time of each event for Wind, ACE and STEREO-A/B. Then, we set an interval of 2 hours, $|t_{\mathrm{Wind}}-t_{\mathrm{s/c}}|<2$ h, as a condition to consider that two events are the same (where the subscript ``s/c'' represents ACE or one of the STEREOs). We therefore end up keeping only the Wind events and removing from the list the same events observed by ACE and STEREOs. In total, we found 235 events that met this condition and were discarded from the overall 1 AU shock list. 

In Figure \ref{shocks1AU_HistSW}, we report in the first row all the IP shocks, in the second row all the IP shocks with a clearly identified magnetic ejecta in the downstream region of the shock, and in the third row IP shocks with SIRs detected. The total number for each category is indicated in the title of each row and the median value of each distribution is indicated as well. We also report in Table \ref{table_ipshocks_icmes} the most probable value and the median value for each distribution, as well as the median values obtained in \cite{kilpua2015properties} for shocks detected with a following ICME and shocks with a SIR detected. 

Overall the distributions of Figure \ref{shocks1AU_HistSW} follow the same tendencies as reported for Figure \ref{Allshocks1AU_HistSW}, with similarities to the results found in \cite{janvier2014mean} with fewer shocks. However, with the higher number of cases studied here, we also find some differences when comparing the second and third rows. For all the parameters, we find that the spread of the distributions is larger for shocks with a detected ICME, with a tail that is longer towards the higher values of each parameter. This is probably due the fact that ICME/shocks have a longer tail composed of stronger shocks than SIR/shocks since the observations made with these spacecraft and even more with Ulysses, showed that the fast coronal hole wind (which drives SIRs) has an upper limit of speeds around 800 km s$^{-1}$ while ICME shocks can exceed this, with occasional events exceeding 1000 km s$^{-1}$.
This implies that the width of the parameters distribution is smaller in the case of the shocks with a detected SIR.

\begin{table}
		\centering
		\begin{adjustbox}{width=1\textwidth}
			\begin{tabular}{cc|cccccc|cccccc|cccccc}\hline
    Data sets & N &\multicolumn{6}{c|}{Most probable value} & \multicolumn{6}{c}{Median value} & \multicolumn{6}{c}{Standard deviation}  \\
	   & & $B_{\rm d}/B_{\rm u}$ & $\dV$ & $N_{\rm p}^{\rm d}/N_{\rm p}^{\rm u}$ & $T_{\rm p}^{\rm d}/T_{\rm p}^{\rm u}$ & $V_{\rm sh}$ & $\MA$ &
	      $B_{\rm d}/B_{\rm u}$ & $\dV$ & $N_{\rm p}^{\rm d}/N_{\rm p}^{\rm u}$ & $T_{\rm p}^{\rm d}/T_{\rm p}^{\rm u}$ & $V_{\rm sh}$ & $\MA$  & $B_{\rm d}/B_{\rm u}$ & $\dV$ & $N_{\rm p}^{\rm d}/N_{\rm p}^{\rm u}$ & $T_{\rm p}^{\rm d}/T_{\rm p}^{\rm u}$ & $V_{\rm sh}$ & $\MA$ \\
			\hline
			All shocks & 720 & 1.54&  30.5&  1.76&  1.54&  418&  1.3 & 1.9& 52.0& 2.0& 1.9& 450 & 1.5 &0.7&51&1.1&2.7&146&0.8 \\ \hline		
			ICME/shocks & 283 & 1.57&  45.6&  2.2&  1.37&  467&  1.2 & 2& 70.6& 2.2& 2.3& 488 & 1.7 &0.8&61&1.2&3.3&178&1.1 \\ \hline
			
			\cite{kilpua2015properties} with ICME & 351 & -&  -&  -&  -&  -&  - & 2& -& 2.2& -& 482& 2.1 &-&-&-&-&-&- \\ \hline
			
			SIR/shocks & 94 & 1.5&  31.8&  1.8&  1.3&  412&  1.2 & 1.8& 41.7& 2& 1.8& 402 & 1.4 &0.6&25.7&1&1.3&78&0.4 \\ \hline
			
			\cite{kilpua2015properties} with SIR & 131 & -&  -&  -&  -&  -&  - & 1.7 & -& 1.9& -& 415 & 1.8 &-&-&-&-&-&- \\ \hline
			

			shocks w/o detected ICME/SIR & 343 & 1.54 &  31.8 &  1.4&  1.37&  413&  1.3 & 1.7& 43.9& 1.8& 1.7& 441 & 1.4 &0.5&41&1&2.3&118&0.6 \\ \hline		
\end{tabular} 
\end{adjustbox}
		\caption{Most probable values, medians and standard deviations of the distributions of\\ parameter defined in Table \ref{table_icmes_1au} caption for IP shocks detected at 1 AU, with a detected ICME\\ and a detected SIR, compared with the results of \cite{kilpua2015properties} of IP shocks with a\\ detected ICME and SIR, respectively. \\
		} 
		
\label{table_ipshocks_icmes}
\end{table}

Table \ref{table_ipshocks_icmes} shows that our values are similar to those reported in \cite{kilpua2015properties} for shocks detected with a following ICME. The magnetic field, proton density ratio and speed shock are the same.  However, the Mach number is higher (2.1) for \cite{kilpua2015properties} compared to our value of 1.7. 

In the case of shocks with detected SIRs, the values are similar, with the exception of the shock speed and Mach number which are higher for \cite{kilpua2015properties}. The standard deviations do not present important changes for each set of data (they are similar to the values given in Section \ref{Sec:Comparing1au}). 
Furthermore, \cite{kilpua2015properties} found that CME-driven shocks are on average slightly stronger and faster, and they show broader distributions of shock parameters than the shocks driven by SIRs. We found the same result in our sample.

The similarity of the distribution shapes, as well as the proximity in the values of the maximum values for all the parameters indicate that the most typical shocks with and without a detected ICME are similar, therefore confirming the conclusion from \cite{janvier2014mean} that shocks at 1 AU are most likely to be ICME-driven, while the detection of a magnetic ejecta is not necessarily made in situ. Indeed, the  results derived by \cite{richardson2018solar} and \cite{lai2012radial} show that the shocks driven by ICMEs often form well inside 1 AU. Furthermore, the rate of shocks associated with ICMEs only changes slightly with solar distance, increasing from 0.72 to 1 AU, then decreasing slightly beyond 1 AU \citep{jian2008stream}.
In contrast, the occurrence of shocks associated with SIRs increases with heliocentric distance. They are less frequent within 0.5 AU of the Sun (at a low rate of around one shock every 200 days) than at 0.5–1 AU (around one shock per 100 days to the Sun), and by 3–5 AU, over 90\% of interaction regions were found to have forward shocks and 75\% have reverse shocks, far higher rates than found at 1 AU or at Helios \citep{richardson2018solar}. Indeed, SIR-driven shocks were reported to start forming at 0.4 AU \citep{lai2012radial}, and the shock association rate with SIRs increases rapidly, from 3\% to 91\%, as they evolve from 0.72 to 5.3 AU \citep{jian2008stream}.

\subsection{Geometrical Properties of the Shocks}
\label{Sec:Geometrical}
  
In the previous studies by \cite{janvier2014mean} and \cite{janvier2015comparing}, the generic shape of the shock front of ICMEs was characterized with a statistical study of shock orientations, from in situ spacecraft data at L1. To do so, the authors introduced the location angle $\lambda$ defined as the angle between the shock normal $\mathbf{\hat{n}}_{\rm shock}$ and the radial direction -$\mathbf{\hat{x}}_{\rm GSE}$. The location angle $\lambda$ can be directly deduced from the components of the normal vector $\mathbf{\hat{n}}_{\rm shock}$ as:
     \begin{equation}
     \tan\lambda = \sqrt{n^{2}_{y} + n^{2}_{z}} \,/\, n_{x}.
     \label{EqLambda}
     \end{equation} 

\begin{figure} [!t]
	\begin{center} 
		\includegraphics[width=11.5cm]{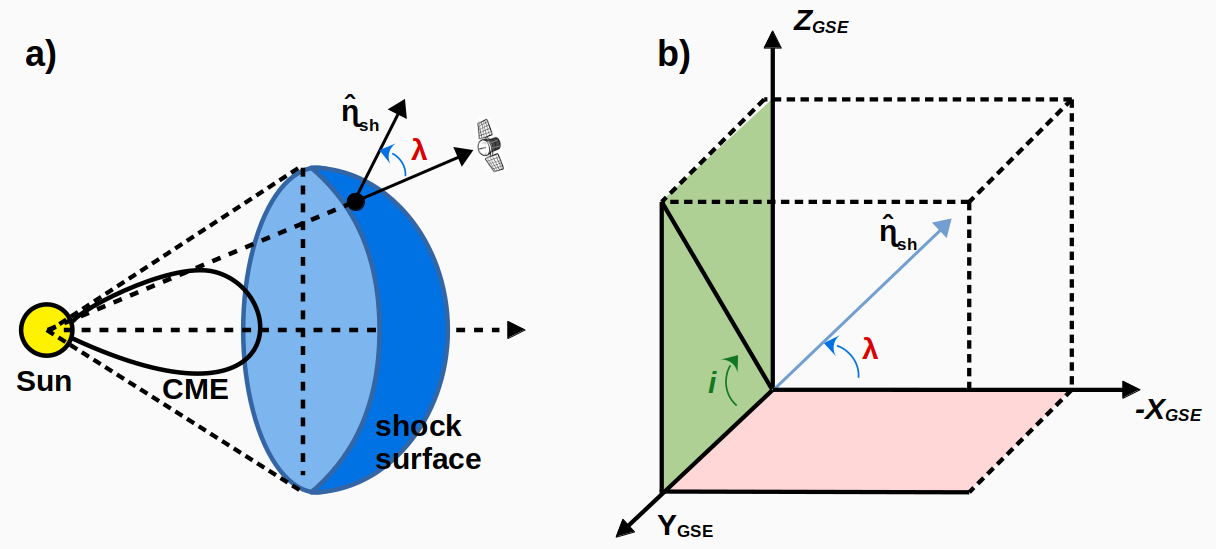}
	\end{center}
\caption{Illustration of the generic shape of the shock front represented by geometric values in GSE coordinates: \textbf{a)} the shock surface is represented here as driven by an ICME. \textbf{b)} The location angle $\lambda$ is the angle between the shock normal vector $\mathbf{\hat{n}}_{\rm shock}$ and the radial direction -$\mathbf{\hat{x}}_{\rm GSE}$, and the inclination angle $i$ measures the inclination between the projected vector $\mathbf{\hat{n}}_{\rm shock,yz}$ and the direction $\mathbf{\hat{y}}$.}
\label{shock_angles_sketch}
\end{figure}

For ICME shocks, this parameter is linked to the relative position of the spacecraft crossing the shock. 
If the location angle $\lambda \sim 0^{\circ}$, then the spacecraft has crossed the IP shock close to its apex, while an increasing value of $|\lambda|$ up to $90^{\circ}$ means that the spacecraft is crossing away from the ``nose'' of the shock (see Figure \ref{shock_angles_sketch}a). 

Through a comparison between the observed distribution of the shock orientations and the synthetic distribution of the parameter $\lambda$, it is possible to deduce the mean shape of the shock front. 
The distribution of the location angle can then be used to infer the general structure of the shock in front of ICMEs \citep[see e.g.][]{janvier2014mean}. Furthermore, the location angle can be used to study the distribution of the shock properties along the shock, from the nose to the wings \citep[see e.g.][]{demoulin2016quantitative}. 

\begin{figure} [!t]
	\begin{center} 
		\includegraphics[scale=0.33]{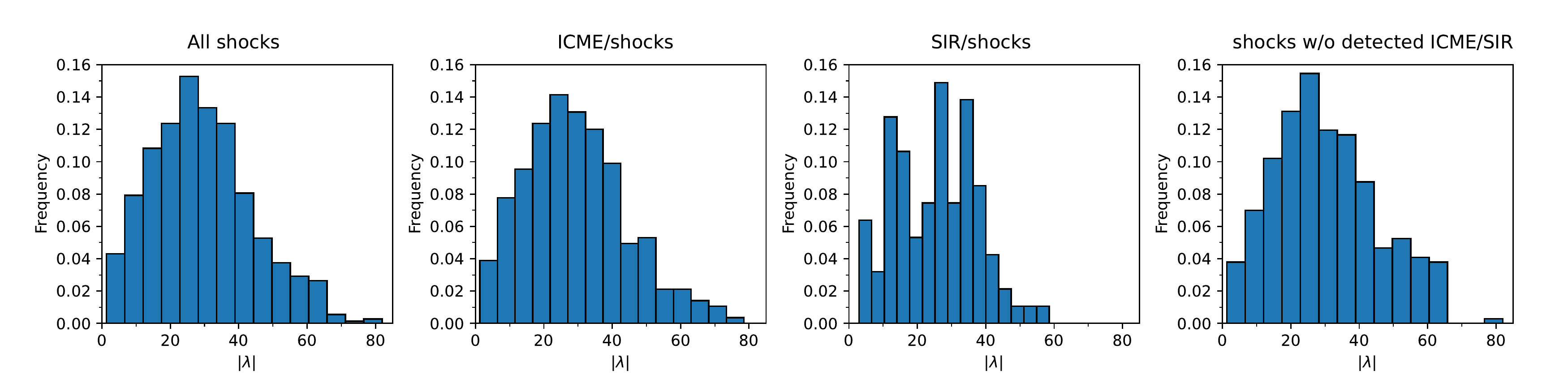}
	\end{center}
	\caption{Distributions of the location angle $\lambda$ using observations at 1 AU. From left to right:
all shocks, the next panel corresponds to all shocks with an ICME detected,
the third panel the shocks with detected SIRs, and the last panel shocks without detected ICMEs/SIRs.}
	\label{lambdaDistributions}
\end{figure}

Figure \ref{lambdaDistributions} shows the distribution of the $\lambda$ angle for shocks detected at 1 AU, for each category. We found similar distributions for shocks detected with ICMEs and those with non-detected ICMEs. Both distributions have similar shape and the same trend, with an abrupt increase at low $|\lambda |$ values and a similar peak location. A slight difference is for the tails. 
The distribution of shocks with detected SIRs indicate that there are less shocks associated with SIRs for large $|\lambda |$ values. 

A second parameter is the inclination angle $i$ on the ecliptic, introduced to quantify the inclination between the projected vector $\mathbf{\hat{n}}_{\rm shock,yz}$ (orthogonal to -$\mathbf{\hat{x}}_{\rm GSE}$) and the direction $\mathbf{\hat{y}}$ (defining $i = 0$ in GSE coordinates), directed towards the solar east, where $i$ ranges from $-180^{\circ}$ to $180^{\circ}$ (see Figure \ref{shock_angles_sketch}b). 

With the coordinate system for spacecraft near Earth, the $x$-axis is toward the Sun and the z axis is directed northward, while y axis complete the orthonormal direct base. We define the value of \textit{i} as:
  \begin{equation}
  i = \atantwo (\vec{n}_{z},\vec{n}_{y}) \,.
  \end{equation}

In the RTN system, R is radially outward from Sun. T is roughly along the planetary orbital direction; N is northward and the RN plane contains the solar rotation axis. Keeping the solar east for the origin of $i$:
  \begin{equation}
  i = \atantwo (\vec{n}_{N},-\vec{n}_{T}) \,.
  \end{equation}

\begin{figure} [!t]
	\begin{center} 
		\includegraphics[scale=0.6]{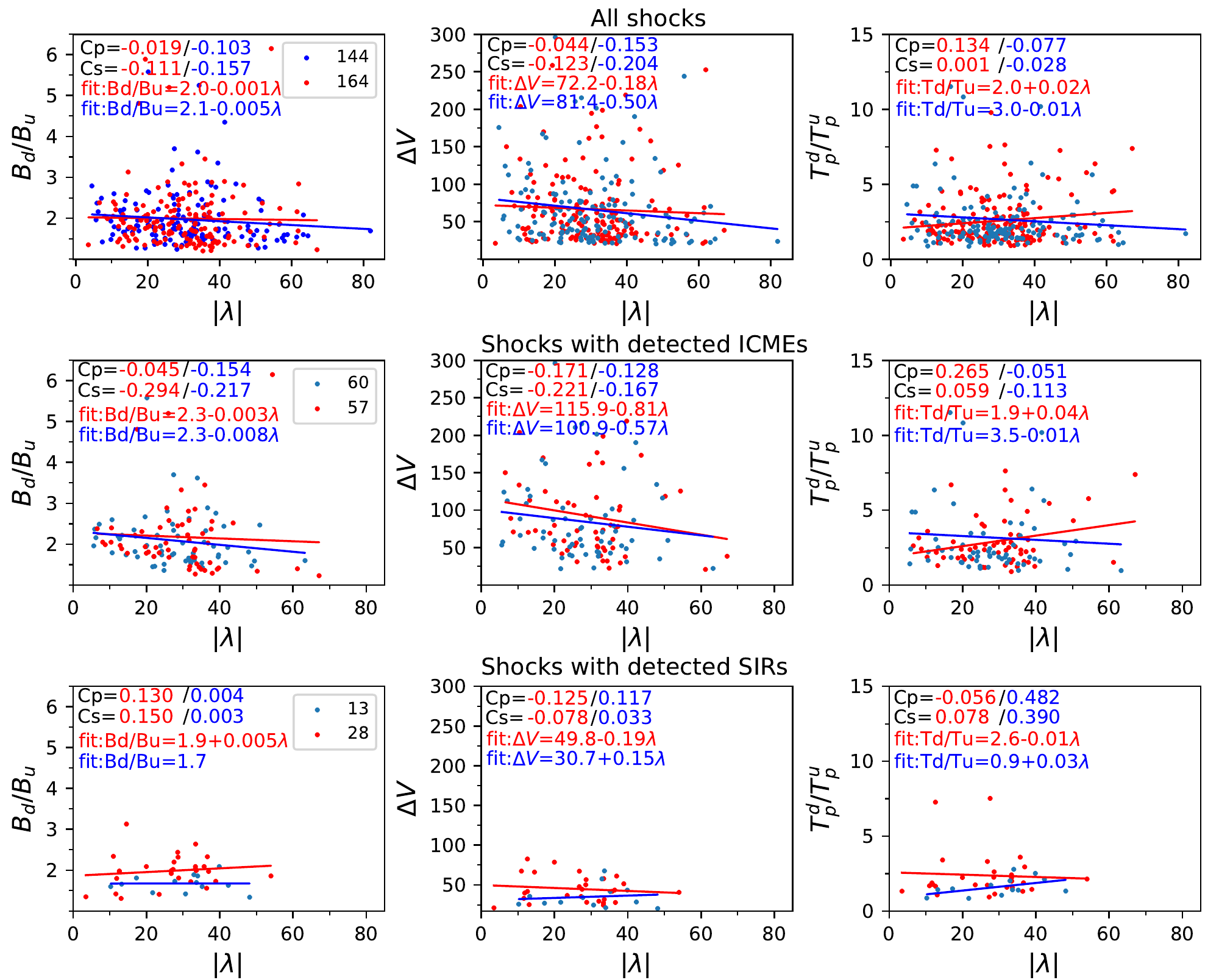}
	\end{center}
	\caption{Correlations between $\lambda$ and the shock parameters for different datasets at 1 AU. The first row corresponds to all IP shocks, the second row to shocks associated with an ICME, and the third row to shocks with detected SIR. In blue are shocks with their normal mostly directed in the solar east direction (with $i$ between $[320^{\circ}, 40^{\circ}]$ with GSE coordinates), and in red in the west direction ($i$ between $[140^{\circ}, 220^{\circ}]$). The Pearson, $c_{\rm p}$, and Spearman, $c_{\rm s}$, correlation coefficients are given, as well as the fit function. }
	\label{1AU_HistSW_Corr}
\end{figure}

Figure \ref{1AU_HistSW_Corr} shows the correlations between the $\lambda$ angle and the shock parameters (magnetic-field ratio, speed jump and proton-temperature ratio) at 1 AU for all the IP shocks in the first row, shocks with a detected ICME in the second row, and shocks with a detected SIR in the third row. The Pearson and Spearman coefficients ($C_{\rm p}$ and $C_{\rm s}$) are given for each distribution as well the linear fitting function. The red/blue codes east/west directions. We analyze below first the red/blue results together before investigating a possible east/west asymmetry. 

We examined possible tendencies between the shock parameters with the location angle $\lambda$ along the shock structure. For the whole data set, the absolute values of the correlation coefficients show that none of the shock parameters, apart $V_{\rm sh}$, is correlated with the location $\lambda$ angle. 
As in \cite{janvier2014mean}, the shock speed, $V_{\rm sh}$, is related to its local radial velocity, $V_{\rho}$, away from the Sun, with the expression $V_{\rm sh} = V_{\rho}\cos(\lambda)$. In the simplest case of a self-similar expansion of an ICME propagating through the solar wind, the outward velocity of the ICME is expected to be radial. In this context, we expect a $\cos(\lambda)$-dependence of $V_{\rm sh}$, where $V_{\rho}$ is non-dependent on $\lambda$ with shocks. So the weak dependence of $V_{\rm sh}$ is a projection effect.

We investigate a possible asymmetry between the shape of the ICME shock, related with the interaction with the Parker spiral, so a possible  east/west asymmetry. The east/west direction is defined with the solar convention.
For that, we select from the whole data set two ranges of $i$: in blue color for those shocks where the angle $i$ is between $[320^{\circ}, 40^{\circ}]$ in GSE coordinates, corresponding to the east direction ($\pm 40^{\circ}$), and in red color for those shocks with $i$ between $[140^{\circ}, 220^{\circ}]$, corresponding to the west direction. We found no significant east-west asymmetry (only three parameters are shown in Figure \ref{1AU_HistSW_Corr}, while similar results are obtained for the other plasma parameters as the ones included in Table~\ref{table_ipshocks_icmes}).

In summary, having found no relevant correlations between the shocks parameters and $\lambda$, we confirm with a bigger sample of IP shocks the results previously found by \cite{janvier2014mean}. The shock shape, deduced from $\lambda$ distribution, does not depend on any parameter of the shocks, i.e., statistically, the shock has a comparable shape regardless of its ICME driver. 

\section{Statistical Analysis of Shock Properties at Different Heliospheric Distances}
\label{Sec:distances}

In the following section, we analyze the distributions of IP shocks properties and their possible evolution at different heliocentric distances using Helios-1/2, 1 AU spacecraft and Ulysses data 
for all the IP shocks detected at different intervals of distances from 0.3 to 5.5 AU.

\begin{sidewaysfigure}
 	\begin{center} 
 		\includegraphics[scale=0.4]{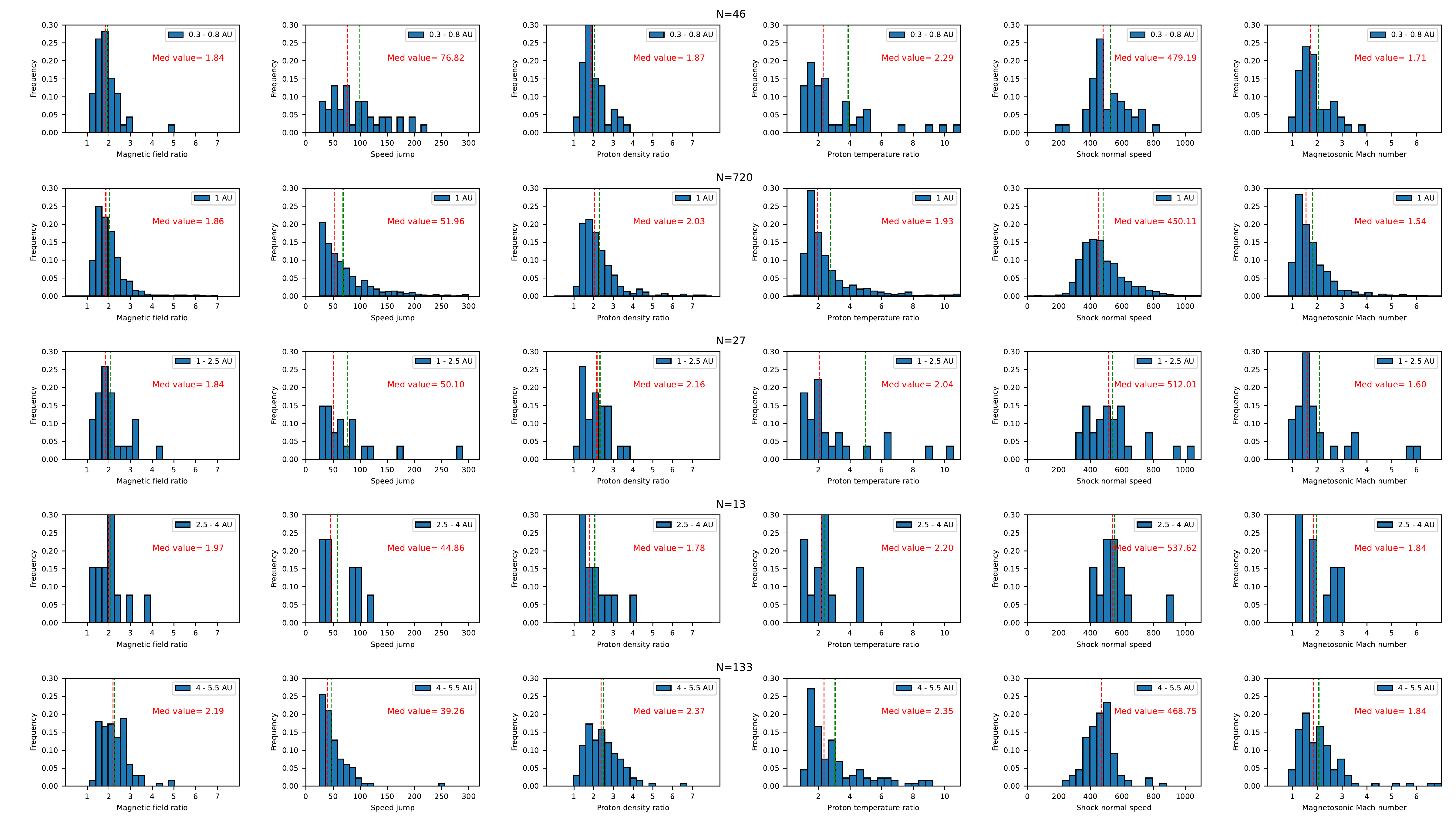}
 	\end{center}
 	\caption{Distributions for all shocks, at different heliocentric intervals, of the magnetic-field ratio, the speed jump, the proton-density ratio, the proton-temperature ratio, the shock normal speed and the magnetosonic mach number. From top to bottom, the rows are for all the shocks reported at [0.3,0.8] by Helios missions, then by 1 AU missions. And finally, by Ulysses spacecraft at three different distances intervals, [1,2.5], [2.5,4] and [4,5.5] AU, and at low latitudes $\pm 30^{\circ}$ degrees, as well. Green and red dashed-lines indicate the mean and median value, respectively.}
 		\label{shocksHelUlyss_HistSW}
\end{sidewaysfigure}

In Figure \ref{shocksHelUlyss_HistSW}, we present the distributions of the shock properties reported at [0.3,0.8] AU by the Helios missions, then at 1 AU and by Ulysses spacecraft at three different distances intervals, [1,2.5], [2.5,4] and [4,5.5] AU, and at low latitudes $\pm 30^{\circ}$ degrees, as well. 
For the cases at [0.3, 0.8] AU, [1,2.5] and [2.5,4] AU the histograms present irregular shapes due to the low number of events. 

\begin{figure}[!t]
	\begin{center} 
		\includegraphics[scale=0.68]{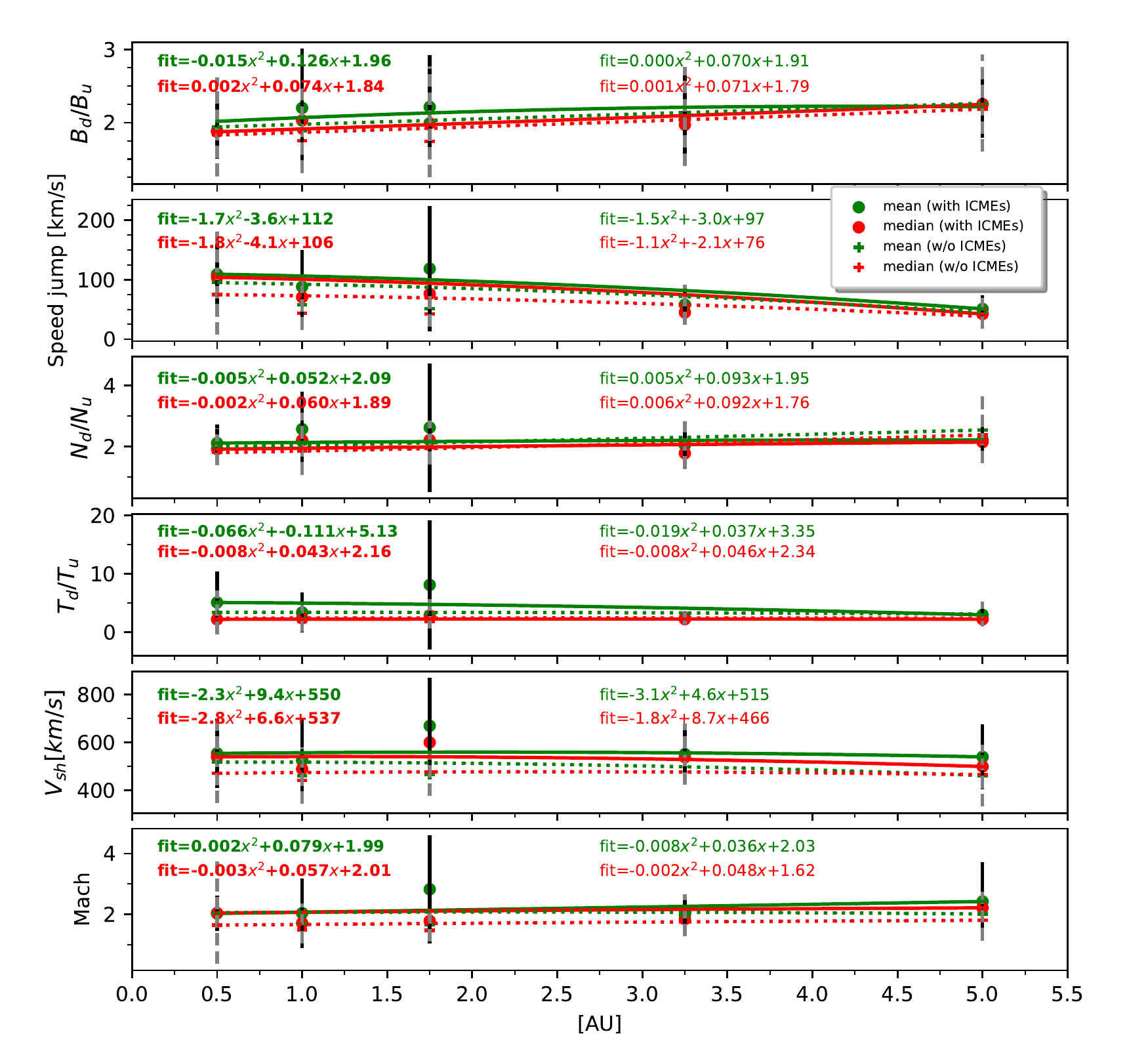}
	\end{center}
	\caption{Variation of mean (green line) and median (red line) values of the shock ratio properties, from 0.3 to 5.5 AU, for shocks with ICMEs (solid line) and shocks without ICMEs (excluding the SIR shocks) detected (dotted-line). In each panel are indicated the general trend of the mean and median values by using a quadratic regression, and the standard deviation of each point indicated as error bars. 
	}
	\label{fittPlot}
\end{figure} 

\subsection{Evolution of Shock Properties with Solar Distance}
\label{Sec:Evolution_properties}

The distributions of the shock parameters are shown as a function of the solar distance in Figure \ref{shocksHelUlyss_HistSW}. In order to outline the radial variation of the mean and median values of the shock properties, we have fitted these quantities by using a quadratic regression (Figure \ref{fittPlot}). We separate shocks with an ICME, solid-line, and shocks without an ICME (excluding the SIR shocks), dotted-line. For both data sets, the green symbols and lines indicates the mean values, and the red symbols and lines the median values.

The colored symbols in Figure~\ref{fittPlot} show more deviations of the mean than the median compared to the global change with distance outlined by the  quadratic regressions. Indeed,  Figure~\ref{shocks1AU_HistSW} show distributions with elongated tails for large values. These tails have a low number of cases so they are more affected by statistical noise than the distribution core.   This result at 1 AU extends to other distances (Figure \ref{shocksHelUlyss_HistSW}).   
The mean is sensitive to the presence of large values in the tail, while the median is more robust to these outsiders, then the results with the median are more robust to the statistical fluctuations, so medians are more trustable.

With the quadratic fit, which has only three free parameters, the results, obtained in five  distance intervals, are coupled together. This implies that the fitted polynomia are less sensitive to statistical fluctuations than within individual intervals.  In particular, the fits of the means and of the medians are nearby with all the parameters (Figure~\ref{fittPlot}). Then, while the number of shocks is limited, especially with Ulysses data, the quadratic fits are expected to provide robust enough results. 

While the fits are quadratic, the results for all parameters are almost a linear dependence with distance.  The largest curvature is obtained with the speed jump $\dV$, but within the error bars this curvature is likely not trustable.  The field ratio, $B_{\rm d}/B_{\rm u}$, has the largest increase with distance. This increase is still relatively modest, about a factor 1.17 from 0.5 to 5 AU. This means that $B_{\rm up}$ is decreasing faster than $B_{\rm down}$ with the distance. This may be due to the strong decrease of the magnetic field in the ambient solar wind with distance from the Sun, as well as the greater presence of rarefaction regions between SIRs reducing the background field at further distances. Both the speed jump $\dV$ and the shock speed have a moderate decrease by a factor 0.45 and 0.89 from 0.5 to 5 AU, respectively. Other parameters remain constant within the error bars. We also investigated the radial variations of the distributions of IP shocks properties with a detected SIR (not shown here). However, we found no significant evolution with distance probably due to the few events detected.

In summary, the shock properties have a weak evolution with heliocentric distances. This agrees with results derived by \cite{lai2012radial} of radial variation of IP shocks from 0.2 to 1 AU.  This weak evolution of the shock parameters contrast with the intrinsic change of some parameters, like magnetic-field strength, plasma density and temperature, since they change by a factor 10 to 100 between 0.5 and 5 AU.

We also investigated the correlations between the location angle $\lambda$ and shock parameters, detected from 0.3 to 5.5 AU (not shown).  Similarly to 1 AU results (Figure \ref{1AU_HistSW_Corr}), we find no relevant correlation between any shock parameters, apart $V_{\rm sh}$, and $\lambda$. 
The slight dependency between the shock speed ($V_{\rm sh}$) with $\lambda$ is shown by both correlation coefficients $|C_{\rm p}|$ and $|C_{\rm s}|$ greater than 0.3. They increase slightly with solar distance since $|C_{\rm p}|$ and $|C_{\rm s}|$ are between 0.3 and 0.5, between 2 and 5 AU. We interpret this dependence as at 1 AU (Section \ref{Sec:WithWithoutICMEs}).



\begin{figure} [!t]
	\begin{center} 
		\includegraphics[width=12cm]{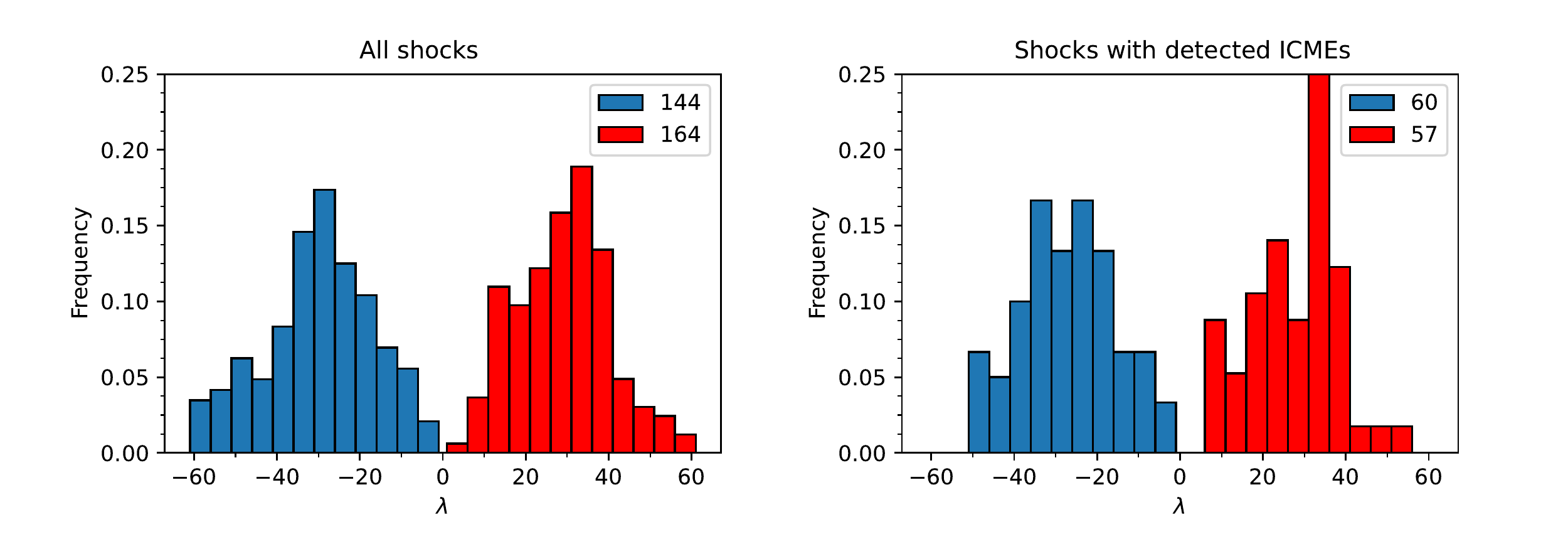}
	\end{center}
	\caption{Distributions of the location angle $\lambda$ using observations at 1 AU with on the left all shocks reported at 1 AU, and on the right all shocks with an ICME detected behind. 
The shock normals directed eastward, with $i$ between $[320^{\circ}, 40^{\circ}]$, are set to artificial negative $\lambda$ values and are shown in blue. The shock normals directed westward, $i$ between $[140^{\circ}, 220^{\circ}]$ are set to positive $\lambda$ values and are shown in red. 
	}
	\label{1AUToLamb}
\end{figure} 


\begin{figure} [!t]
	\begin{center} 
		\includegraphics[width=12cm]{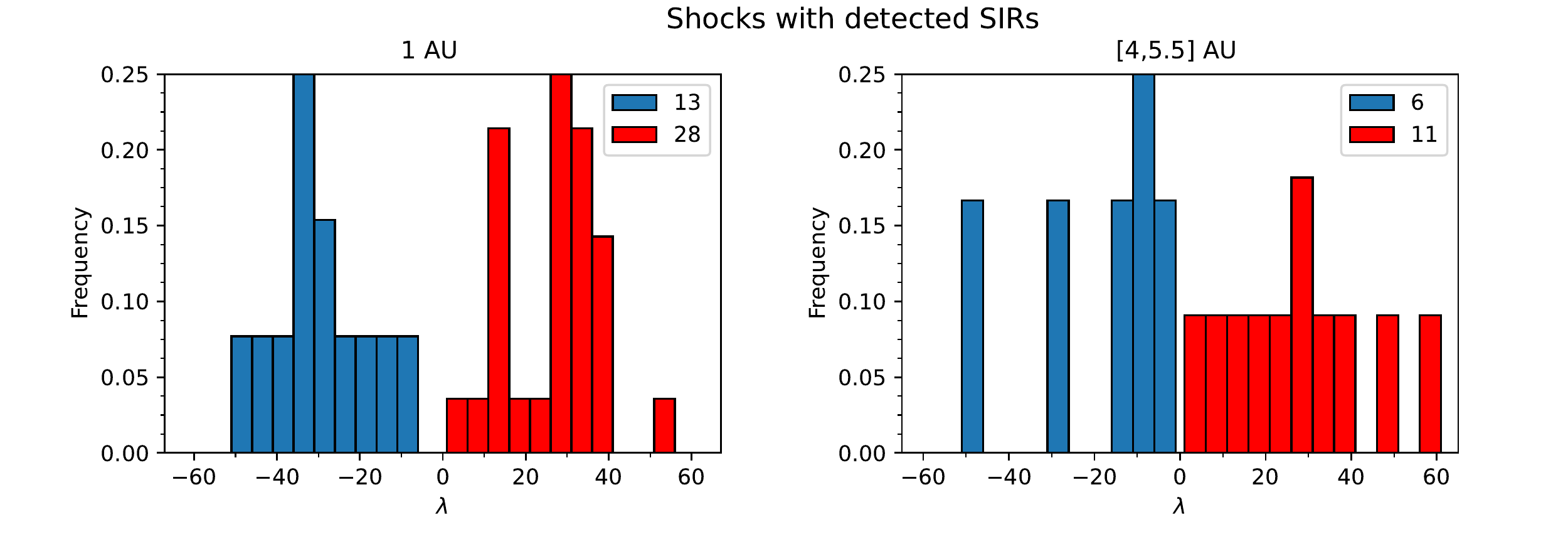}
	\end{center}
\caption{ Distributions of the location angle $\lambda$ for shocks with detected SIRs at 1 AU and [4,5.5] AU, respectively. The blue color is for those shocks where the angle $i$ is between $[320^{\circ}, 40^{\circ}]$ (shock normal towards east direction, $\lambda$ set artificially negative), and the red color for those shocks with $i$ between $[140^{\circ}, 220^{\circ}]$ (shock normal towards west direction, $\lambda$ positive).} 	\label{lambdaDistributionsHelUly02}
\end{figure} 

\subsection{Distribution of the Location Angle $\lambda$} 
\label{Sec:DistLambda}

Considering the number of shocks detected at 1 AU in our study (720 events), we are extending the shocks database from \cite{janvier2014mean}, in order to obtain more reliable results. As well as, since most of the shocks, detected from Ulysses, were between 4 and 5.5 AU and at low latitudes $\pm 30^{\circ}$, in the present section we will focus on this interval. To further investigate a possible asymmetry between the shape of the ICME/shock, we have separated eastward and westward values as an artificial construction with the east/west solar direction of the shock normal. In Figure \ref{1AUToLamb} the $\lambda$ distributions for those shocks where the angle $i$ is between $[320^{\circ}, 40^{\circ}]$, corresponding to the east (in blue) solar direction ($\pm 40^{\circ}$), and in red color for those shocks with $i$ between $[140^{\circ}, 220^{\circ}]$, corresponding to the west solar direction ($\pm 40^{\circ}$).  The $\lambda$ values are separated in negative (red) and positive (blue) to investigate the shape asymmetry of the ICME/shock.

In both directions, east and west, we found similar distributions for all shocks detected at 1 AU (Figure~\ref{1AUToLamb}, left panel) so there is no indication of a bias induced by the Parker spiral. 
The low number of ICME shock (middle panel) detected for small values of $|\lambda|$ values can be explained by the small extension of the shock surface near the apex \citep{janvier2014mean}. So that, there is less chance of crossing the shock near its apex that at other parts of the structure. Next, for large values of $|\lambda|$, between $50^{\circ}$ and $60^{\circ}$, there are more shocks without than with ICME behind (not shown).
Indeed, when a spacecraft crosses the structures farther from the apex, the encounter with the following ICME is likely too near its boundary to detect the ICME, or even the crossing can be outside the ICME. Then, shocks with large $|\lambda|$ values are expected to be more associated with non-detected ICMEs.

In Figure \ref{lambdaDistributionsHelUly02} we plotted the $\lambda$ distributions, with 15 bins, for all shocks detected with SIRs, at 1 AU, and between 4 and 5.5 AU. 
As the number of cases at 1 AU is reduced to about the number observed between 4 and 5.5 AU, the distribution become less regular. 
Despite the limited statistics, one would expect more westward oriented shocks events with SIRs compared to eastward oriented given the spiral geometry of interaction regions in the ideal corotating geometry and indeed this is the case by a factor two at both distances.

From Ulysses data, most shocks have no detected ICME behind, while it is the opposite at and below 1 AU. 
Indeed, with Ulysses data SIRs are the primary drivers of FF shocks (34 \%), while ICMEs represent only 17 \%, so the rest, 49 \% have no detected structure behind. 
Furthermore, from 1 to 5 AU the fraction of shocks with SIRs increase slightly from 21 \% to 34 \%, and the fraction of shocks with ICMEs decreased from 47 \% to 17\%. A possible interpretation is that shocks detected by Ulysses are most likely to be CIR/SIR, as the interaction between fast and slow winds needs time, so distance, to build up in a strong shock. In contrast, ICMEs are typically launched fast from the Sun, then they progressively slow down towards the surrounding solar speed. Then, ICME shocks are expected to be stronger in the inner heliosphere. Then, shocks within 1 AU are expected to be dominantly ICME driven, while those detected by Ulysses are most likely to be CIR/SIR driven.  

\section{Conclusions}
\label{Sec:Conclusion}

The main aim of this paper is to present a statistical study of a large number of shocks in order to understand the evolution of shock properties with heliospheric distance, as well as to study if there is a possible preferential orientation of shocks as a function of the inclination angle $i$ and the location angle $\lambda$.

Our study is based on a series of catalogs reporting parameters associated with ICMEs, IP shocks and SIRs detected at several heliocentric distances, from 0.29 to 0.99 AU with Helios-1/2; near 1 AU using Wind, ACE and STEREO-A/B; and 1.3 to 5.4 AU with Ulysses. We developed a multi-spacecraft analysis of interplanetary shocks at 1 AU in order to investigate spacecraft biases. We present a comparison between the shock normal vectors, for those events detected at 1 AU by Wind, reported in two different databases: from the Heliospheric Shock Database and CfA Interplanetary Shock Database. From this comparison, we found a strong correlation (above 0.8) between the normal components from the HSD and the shock normal methods reported in CfA database. This result gave us confidence that the Heliospheric Shock Database offers us quite reliable results.

We do not found significant differences, of the shock properties, for spacecraft positioned at 1 AU, as expected.  
We investigated the distribution of all shocks with detected and non-detected ICMEs at 1 AU. Shocks associated with detected ICME showed higher mean values of the downstream to upstream ratios of the shock for the magnetic-field strength, density and temperature than the shocks without a detected ICME. The obtained result that faster and stronger shocks are associated with ICMEs, is in agreement with several related studies  \citep[][and references therein]{janvier2019generic}. 
Moreover, the similarity of the shock distributions reinforces the conclusion from \citet{janvier2014mean} that shocks at 1 AU are most likely to be ICME-driven, while the detection of a magnetic ejecta is not necessarily made in situ (the shock being more extended ortho-radially than the magnetic ejecta). 
Next, we examined the relation between the shock properties with the location angle $\lambda$, which defines, for a given shock shape, the location of the spacecraft crossing the ICME/shock structure. In accordance with the results found by \citet{janvier2014mean}, with a bigger sample of shocks, we do not found relation between the shock parameters with the location angle, so that the shock front has statistically uniform properties.

We next analyzed the shock properties by interval of latitude with Ulysses. We found that most shocks were detected at low latitudes within $\pm 30^{\circ}$ and beyond 4 AU. 
We also studied the distributions of shock properties and their possible evolution with distance from the Sun with Helios-1/2, 1 AU missions and Ulysses (latitude below $30^{\circ}$) spacecraft. We found at most 15 \% variation with solar distance (0.5 - 5 AU) in the parameter ratios at the shock. This is in contrast with a typical factor 10 or 100, depending on the parameter, of the corresponding plasma parameters in the solar wind and in ICMEs.  
We found no relevant variations in the $\lambda$ distributions with heliocentric distance, as well as preferential direction of {$\mathbf{\hat{n}}_{\rm shock}$} around the Sun apex line, by separating east/west oriented ICME/shocks. In contrast, SIR/shocks are about twice more numerous with a westward than eastward shock normal orientation.

We have studied the variation, spatial and temporal, of the distribution of shock properties from several space missions. 
Our statistical analysis developed in this study can be applied to understand the evolution of ICME/shocks with distance in the inner Heliosphere. We need a consistent set of in-situ observations to obtain more reliable results. In that sense, new missions such as Parker Solar Probe \citep{case2020solar} and Solar Orbiter \citep{muller2020solar} will be of great help to study the evolution and propagation of ICMEs and interplanetary shocks at closer solar distances. \\

\textbf{Acknowledgments} 
Carlos Pérez-Alanis and Ernesto Aguilar-Rodr\'{i}guez acknowledge support from DGAPA/PAPIIT project IN103821.
We recognize the collaborative and open nature of knowledge creation and dissemination, under the control of the academic community as expressed by Camille No\^{u}s at www.cogitamus.fr/indexen.html.

\appendix

\section{How Well is the Shock Normal Determined?} \label{Appendix:A}

In the present study, we used the IP shock catalog provided by the Heliospheric Shock Database (HSD) maintained at the University of Helsinki, in order to calculate the location angle lambda $\lambda$ and compare it using different datasets of shock normal to check whether it is a quantity that is well-defined. To construct this database, two methods to identify shocks have been used: (1) through a visual inspection of solar-wind plasma and magnetic-field parameters, simultaneously looking for sudden jumps in the plasma and magnetic field parameters. These jumps should satisfy the shock wave signatures to be considered as forward (FF) or fast reverse shock (FR), and (2) using an automated shock detection machine-learning algorithm, called IPSVM (InterPlanetary Support Vector Machine) \citep{Isavnin2007}. 

According to the documentation of the HSD, the shock normal vector ({$\mathbf{\hat{n}}_{\rm shock}$}) is calculated using the mixed-mode method (MD3). This method follows the conditions that the cross product of the upstream and downstream magnetic fields should be perpendicular to the shock normal provided there are no gaps in the velocity components.
In the case of a data gap in the velocity components, the normal vector is calculated using the magnetic-field coplanarity, which is also the most widely used method. It relies on the coplanarity theorem and is based on the idea that the normal vector to a planar surface can be determined if two vectors lie within this surface. 

For compressible shocks, the magnetic field on both sides of the shock and the shock normal all lie in the same plane. Thus there is a variety of vectors which lie in the shock plane. These include the change in magnetic field, the cross-product of the upstream and downstream magnetic field (which is perpendicular to the coplanarity 
plane), and the cross-product between the upstream or downstream magnetic field and their difference with the change in bulk velocity (\citeauthor*{paschmann2000issi}, 2000 and references therein). The magnetic coplanarity normal use the cross-product of the upstream and downstream magnetic field, and although it is easy to apply, it fails for $\theta_{B_{n}}=0^{\circ}$ or $90^{\circ}$. 
Regardless of the method used to determine the shock normal vector and related parameters its determination may have significant errors depending on the upstream and downstream values chosen for each shock. In order to obtain the correct results of the shock parameters, the shock has to be entirely exclude from the upstream and downstream data points and not consider disturbances not related with the shock. However, the upstream and downstream intervals vary depending on the shock and in most cases there is no well defined criteria for choosing these intervals.

We then calculated the angle $\lambda$ from the normal shock vectors applying Equation \ref{EqLambda}.
This database has the advantage of applying the same method to a large set of spacecraft data (see Section \ref{Sec:selection}), and therefore ensures consistency throughout the analysis. However, other authors have also worked on the detection as well as the characterization of the shock parameters, including the normal vector that we use to obtain the location angle. In the following, we therefore compare  different available datasets of shock normal to check whether it is a quantity that is well determined.

\begin{figure} [!t]
	\begin{center} 
		\includegraphics[width=11.5cm]{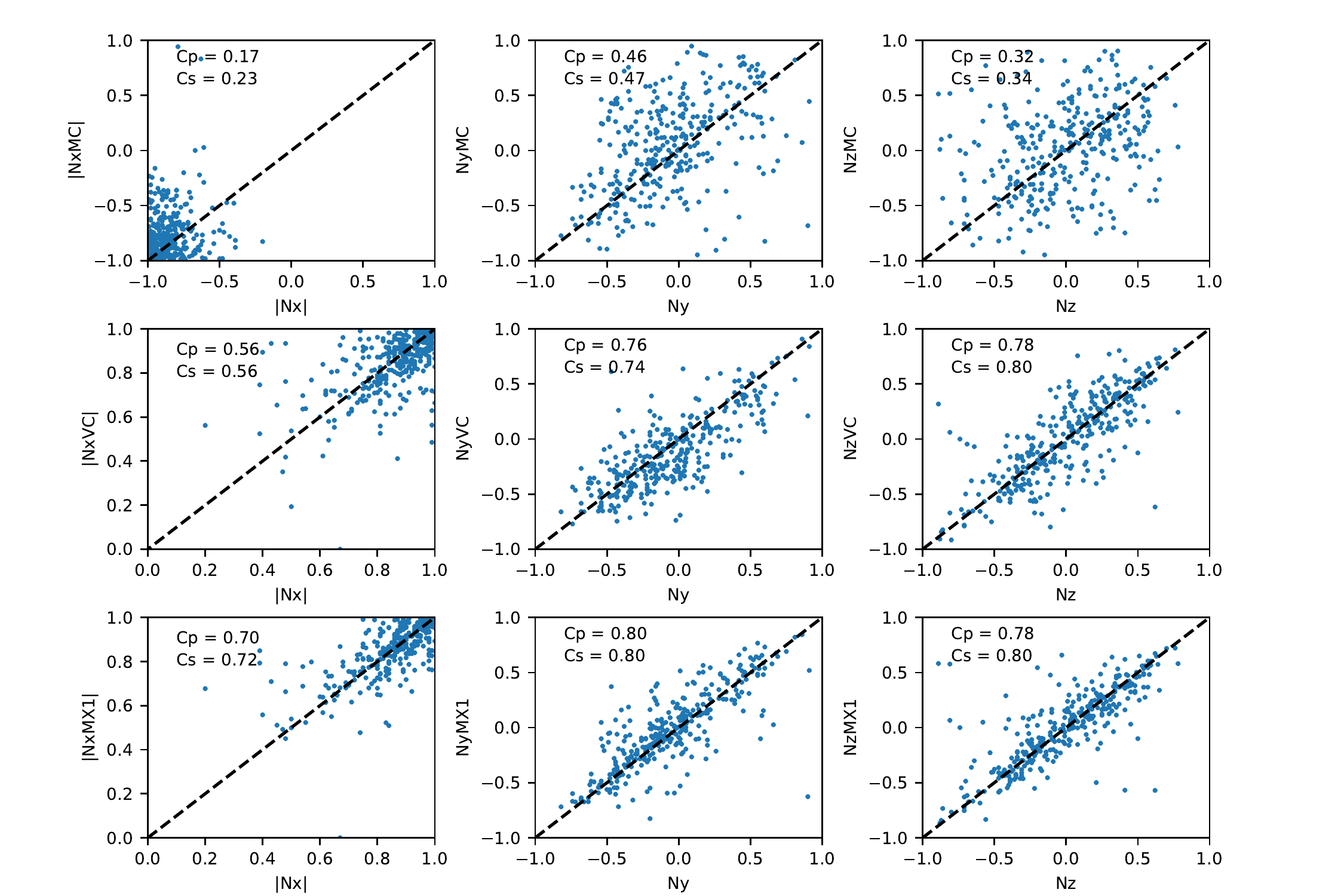}
	\end{center}
	\caption{Correlations between each component of the shock normal, calculated with three methods (the CfA shock database), MC, VC, and MX1, with the mixed mode method (Heliospheric Shock Database) (see Section \ref{Sec:databases}). For each graph, we have indicated the Pearson and Spearman correlation coefficients, $C_{\rm p}$ and $C_{\rm s}$.}
\label{MethComp}
\end{figure}

\begin{figure} [!t]
	\begin{center} 
		\includegraphics[width=8cm]{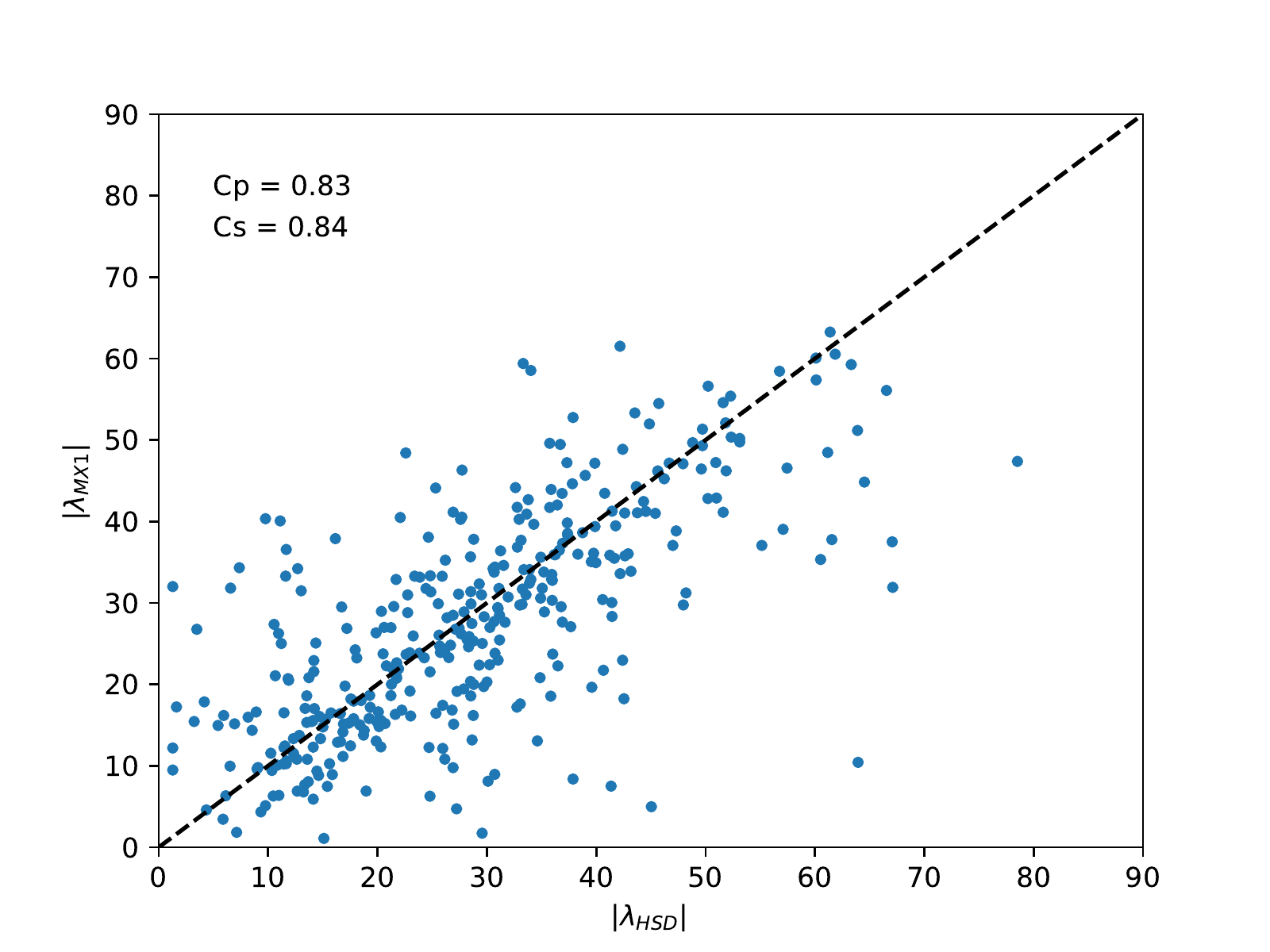}
	\end{center}
	\caption{Correlation between the $\lambda$ derived by the mixed mode method ($\lambda_{\rm HSD}$) with $\lambda$ calculated from the MX1 ($\lambda_{\rm MX1}$). The Pearson and Spearman correlation coefficients have been added in the top left corner.}
	\label{LambdaComp}
\end{figure} 

\subsection*{Shocks at 1 AU}

For shocks detected at 1 AU, the CfA Interplanetary Shock Database\footnote{lweb.cfa.harvard.edu/shocks/} provides IP shocks observed by the Wind spacecraft. We therefore explore in the following the different methods used in this database, and compare the location angle outputs with that of the Heliospheric Shock Database.

In the CfA Interplanetary Shock Database, the events are listed by year with a detailed analysis for each shock detected. Each shock was analyzed {with} the shock normal vectors derived by three methods: magnetic coplanarity (MC), velocity coplanarity (VC) and three mixed mode (MX1).

The velocity coplanarity (VC), not used in the Heliospheric database, offers a good approximation to the normal vector, which is valid at high Mach numbers and for $\theta_{B_{n}}$ near $0^{\circ}$ or $90^{\circ}$ \citep{daly2000analysis}. The three mixed mode method, referred to as MX1 in the following, requires both plasma and field data to calculate the components of the shock normal vector. For this, each component of the upstream, downstream and magnitude of the magnetic field is used. 

Through a comparative analysis, we selected events with 1 hour or less difference between the detection time reported in both catalogs. Most events can be found in both catalogs, so that we ended up with 358 total events, around $69\%$ of the total amount of events reported in the Heliospheric Shock Database.

Figure \ref{MethComp} shows the correlations between each component of the normal vector ($N_{x}$, $N_{y}$, $N_{z}$) from the Heliospheric Shock Database with the normal component of the CfA Interplanetary Shock Database from each method: ($N_{x\rm MC}$, $N_{y\rm MC}$, $N_{z\rm MC}$) for the magnetic coplanarity {method}, ($N_{x\rm VC}$, $N_{y\rm VC}$, $N_{z\rm VC}$) for the velocity coplanarity method and ($N_{x\rm MX1}$, $N_{y\rm MX1}$, $N_{z\rm MX1}$) for the three mixed mode method.

We use the absolute value of the $x$-component to compare positive values. These components are grouped in a small region (Figure \ref{MethComp}, right panels). We find that the MC method shows the lowest correlation with the results from the HSD, as shown in the first row, and the MX1 method shows the best correlations for all components, with Pearson and Spearman correlation around 0.7$\sim$0.8 for all components. 

In Figure \ref{LambdaComp}, we show the correlation between the location angle derived by the shock normal vector from the mixed mode method ($\lambda_{\rm HSD}$ reported in the Heliospheric Shock Database), with the location angle derived by the shock normal vector from the MX1 method ($\lambda_{\rm MX1}$, reported in the CfA Interplanetary Shock Database). Both Pearson and Spearman coefficients ($C_{\rm p}$ and $C_{\rm s}$) indicate a strong correlation between both $\lambda$ values, with 0.83 and 0.84, respectively.

In summary, we do not find significant differences between the $\lambda$ values derived by different shock normal methods. The strongest correlations are obtained, as expected, for the mixed mode method used for both the HSD and the CfA Interplanetary shock database. Since the results are highly correlated, with correlation coefficients above 0.8, this comparison gives us confidence that the Heliospheric Shock Database offers us quite reliable results, while also providing a large coverage of several spacecraft at different heliodistances.

\bibliographystyle{spr-mp-sola}
\bibliography{bio} 

\end{article} 

\end{document}